\documentclass[twocolumn,showpacs,pra]{revtex4}
\usepackage{amssymb}
\usepackage{amsmath}
\usepackage{graphicx}
\usepackage{dcolumn}
\usepackage{bm}
\usepackage{subfigure}
\usepackage{xcolor}

\begin{document}

\title{The optimal approximation of qubit states with limited quantum states}
\author{Li-qiang Zhang$^1$}
\author{Deng-hui Yu$^1$}
\author{Chang-shui Yu$^{1,2}$}
\email{ycs@dlut.edu.cn}
\affiliation{$^1$School of Physics, Dalian University of Technology, Dalian 116024, China }
\affiliation{$^2$DUT-BSU joint institute, Dalian University of Technology, Dalian 116024, China }

\begin{abstract}
Measuring the closest distance between two states is an alternative and significant approach in the resource quantification, which is the core task in the resource theory.  Quite limited progress has been made for this approach even in simple systems due to the various potential complexities.
Here we analytically solve the optimal scheme to find out the closest distance between the objective qubit state and all the possible states convexly    mixed by some limited states, namely, to optimally construct the objective qubit state
using the quantum states within any given state set. In particular, we find
the least number of (not more than four) states within a given set to optimally construct the objective state and also find that any state can be optimally established by at most four quantum states of
the set. The examples in various cases are presented to verify our analytic
solutions further.
\end{abstract}

\maketitle

\emph{Introduction.}-Quantum mechanical intriguing features are the
essential ingredients that distinguish the quantum and the classical worlds,
and at the heart of the quantum information, because they can be exploited
to achieve the quantum information processing tasks and hence serve as the
unsubstitutable physical resources. The resource theory, which
quantitatively characterizes the quantum features in a mathematically
rigorous fashion, has been extensively developed in recent years \cite%
{R1,R2,R3,R4,R5,R6}. There are two ingredients in a well-defined resource
theory: the free states, which do not owe any quantum feature of interest,
and the free operations, which cannot convert the free states to the
resource states (opposite to the free states). Besides, a quantitative
framework for any quantum feature also depends on how to discriminate the
given state from the free states unambiguously \cite%
{E01,E02,D4,D5,C1,C2,C3,F1}.

 The most remarkable example is the quantification of entanglement \cite{E1,E2,E3}, which is essential to find the
least distance between the given state and the separable-state set. Since the convex mixing of free states is still within the free-state set, the entanglement measure can also 
be regarded to find out the least distance between the given state and all potential convex mixings of separable states. Similarly, the quantum discord of a state can also be quantified
by the distance  from the closest convexed mixed local distinguishable
states \cite{D4,D5,D6,D7,D8,D9,D10,D11,D12,D13,D14}. Besides, the widely
studied coherence can be measured with the least distance from all the possible
convex mixings of some given orthonormal basis \cite%
{C1,C2,C3,C4,C6,C7,C8,C10,C11,C12,C13}. In this sense, the essential problem is 
to optimally approximate a given state by the convex mixing of the limited states.

The state approximation via the convex mixing of some limited states has its
much more general significance. Suppose, in some particular experiment, one
can only prepare the limited pure states whose set is denoted by $\left\{
\left\vert \psi _{i}\right\rangle \right\} $. But a general state $\rho $ is
required for some purpose. The question is, what is the best way to
approximating the state $\rho $ by the convex mixing of the states in set $%
\left\{ \left\vert \psi _{i}\right\rangle \right\} $ and how well the state $%
\rho $ can be approximately replaced. Recently, the question has been
addressed for some particular cases, such as approximating the given state
by the eigenstates of several Pauli matrices based on various distance
measures \cite{CC1,CC2,FSM}. A related question was also considered for the
optimal approximation of a desired and unavailable quantum channel $\Phi $
by the convex mixing of a given set of other channels $\{\Psi _{i}\}$ \cite%
{CC3}.

In this paper, we analytically solve the optimal construction of a qubit
state with the convex mixing of pure/mixed states subject to any given quantum state set
\cite{T1,T2,T3,T4,T5}. We find the least number of states in a given set to optimally construct the objective state. In particular, we show that any qubit state can be optimally
constructed by at most four states, no matter how many quantum states are
given. Therefore, the optimal construction of a qubit state with more than
four quantum states can be converted into the case with at most four quantum
states, which is perfectly solved by the analytic and closed solution. In
addition, we present some numerical and analytic examples to
verify/demonstrate our analytic results.

\emph{The scheme.}-Given a quantum state set $S:=\left\{
\rho_{i} ,i=1,2,\cdots ,N\right\} $ and an
expected objective qubit state  $\rho $, the task is to prepare a state $%
\sigma =\chi _{1,2,\cdots ,K}\left( \vec{p}\right)
=\sum\limits_{i=1}^{K}p_{i}\rho_{i}$ by the convex mixing of $K\leq N$ quantum states in
the set $S$ such that $\sigma $ and $\rho $ are as close as possible subject
to the state distance
\begin{equation}
D\left( \rho ,\sigma \right) =\left\Vert \rho -\sigma \right\Vert _{1},
\end{equation}%
where $\left\Vert X\right\Vert _{1}$ is the trace norm with $\left\Vert
X\right\Vert _{1}=\mathrm{Tr}\sqrt{X^{\dagger }X}$. It is obvious that $%
D\left( \rho ,\sigma \right) $  is (joint) convex
and contractive under trace-preserving quantum operation. 
$D\left( \rho,\sigma \right) =0$ for $\rho =\sigma $ and $D\left( \rho ,\sigma \right) =2$
for $\rho \perp \sigma $.  
In addition, it is implied that only $K\leq N$
states could achieve the optimal $\sigma $. For example, if the two
eigenstates of $\rho $ happened to be covered in the set $S$, $\rho $ can be
perfectly constructed by the two ($k=2$) states.

For a qubit state $\rho $, the trace norm between  $\rho $ and $\sigma =\chi _{1,2,\cdots ,K}\left( \vec{p}\right) $ is
given by%
\begin{equation}
D\left( \rho ,\sigma \right) = \left\vert \lambda _{1}\right\vert +\left\vert
\lambda _{2}\right\vert =2\left\vert \lambda _{1}\right\vert ,
\end{equation}%
with $\lambda _{i}$ the eigenvalues of Hermitian matrix$\ \rho -\sigma $. In the Bloch representation,
let $\mathbf{r}_{o}$ denote the Bloch vector of $\rho $ with its elements $%
r_{o\alpha }=$ Tr$(\rho \sigma _{\alpha })$ and $\mathbf{r}_{k}$ ($%
r_{k\alpha }$ is its element) denote the Bloch vector of the $k$th state in $%
S$, where $\sigma _{\alpha }$ are Pauli matrices with $\alpha =x,y,z$. Then
we get%
\begin{equation}
D^{2}\left( \rho ,\sigma \right) =\sum_{\alpha ij}p_{i}p_{j}r_{i\alpha
}r_{j\alpha }-2p_{i}r_{i\alpha }r_{o\alpha }+r_{o\alpha }^{2}.  \label{D1}
\end{equation}%
Therefore, in the scheme, to construct the state $\sigma $ close enough to
the state $\rho $ is equivalent to achieve $\min_{\vec{p}}D^{2}\left( \rho
,\chi _{1,2,\cdots ,N}\left( \vec{p}\right) \right) $.

It is obvious that the considered scheme is essentially an optimization
problem. The Hessian matrix for this problem is defined by $H=\frac{\partial
^{2}D^{2}}{\partial \vec{p}^{2}}$, so one will immediately find that
\begin{equation}
H=\frac{\partial ^{2}D^{2}}{\partial \vec{p}^{2}}=2R^{T}R,  \label{hessian}
\end{equation}%
with $\mathcal{R}=\left( \mathbf{r}_{1},\mathbf{r}_{2},\cdots ,\mathbf{r}%
_{N}\right) $ being a $3\times N$ matrix. Then the optimized function $%
D^{2}\left( \rho ,\sigma \right) $ is convex on $\vec{p}$. It indicates that
our considered scheme is a convex optimization problem \cite{T3}, which
plays the dominant role in the optimal state construction. It allows us to
find the optimal scheme on the boundaries of the constraints if the globally
optimal points are beyond the constraints, which is the cornerstone of our
whole letter.

\emph{The main results.}-Now we can go forward to the optimal scheme by
considering the different $N$, the number of the quantum states in the set $S$,
in a rigorous way. Following these results below, one can directly find out the
best way to approximately preparing a density matrix by any given quantum
states, and the optimal distance between the prepared state and the
objective state.

\textbf{Theorem 1}.- Let the set $S$ contain two states ($N=2$%
). If $0\leq(\mathbf{r}_{o}-\mathbf{r}_{2})^{T}(\mathbf{r}_{1}-\mathbf{r}_{2})\leq\Vert\mathbf{r}_{1}-\mathbf{r}_{2}\Vert_{2}^{2} $, where $\left\Vert \mathbf{r}\right\Vert _{2}=\sqrt{\mathbf{r}^{T}\mathbf{r}}$,
one can construct an optimal state $\chi _{1,2}\left( \vec{p}\right) $ with
the optimal distance given by%
\begin{eqnarray}
D^{2}(\rho,\chi_{1,2}\left( \vec{p}\right))=\Vert\mathbf{r}_{o}-\mathbf{r}_{2}\Vert_{2}^{2}-\frac{\left[  \left( \mathbf{r}_{o}-\mathbf{r}_{2} \right)^{T}\left( \mathbf{r}_{1}-\mathbf{r}_{2} \right) \right] ^{2} }{\Vert\mathbf{r}_{1}-\mathbf{r}_{2}\Vert_{2}^{2}},
\end{eqnarray}%
and the optimal weight $\vec{p}$ given by
\begin{eqnarray}
p_{1}&=&\frac{\left( \mathbf{r}_{o}-\mathbf{r}_{2} \right)^{T}\left( \mathbf{r}_{1}-\mathbf{r}_{2} \right)}{\Vert\mathbf{r}_{1}-\mathbf{r}_{2}\Vert_{2}^{2}} ,  \notag \\
p_{2}&=&1-p_{1}.
\end{eqnarray}%
If $(\mathbf{r}_{o}-\mathbf{r}_{2})^{T}(\mathbf{r}_{1}-\mathbf{r}_{2})<0$, the optimal distance is given by
\begin{eqnarray}
D(\rho ,\chi _{1,2}\left( \vec{p}\right) )=D(\rho ,\chi _{2}\left( \vec{p}\right) )=\Vert\mathbf{r}_{o}-\mathbf{r}_{2}\Vert_{2}.
\end{eqnarray}%
If $(\mathbf{r}_{o}-\mathbf{r}_{2})^{T}(\mathbf{r}_{1}-\mathbf{r}_{2})>\Vert\mathbf{r}_{1}-\mathbf{r}_{2}\Vert_{2}^{2}$, the optimal distance is given by
\begin{eqnarray}
D(\rho ,\chi _{1,2}\left( \vec{p}\right) )=D(\rho ,\chi _{1}\left( \vec{p}\right) )=\Vert\mathbf{r}_{o}-\mathbf{r}_{1}\Vert_{2}.
\end{eqnarray}%
\textbf{Proof}. For $N=2$, Eq. (\ref{D1}) can be rewritten as%
\begin{equation}
D^{2}(\rho ,\chi _{1,2}\left( \vec{p}\right) )=\sum_{ij}^{2}p_{i}p_{j}%
\mathbf{r}_{i}^{T}\mathbf{r}_{j}-2p_{i}\mathbf{r}_{o}^{T}\mathbf{r}_{i}+%
\mathbf{r}_{o}^{T}\mathbf{r}_{o}.  \label{d1}
\end{equation}%
Consider the Lagrangian function%
\begin{eqnarray}
&&L(p_{1},p_{2},\lambda ,\lambda _{1},\lambda _{2})  \\  \label{l1}
&=&D^{2}(\rho ,\chi _{1,2}\left( \vec{p}\right) )-\lambda _{1}p_{1}-\lambda
_{2}p_{2}+\lambda \left( p_{1}+p_{2}-1\right) ,  \notag
\end{eqnarray}%
where $\lambda $ and $\lambda _{i}$ are the Lagrangian multipliers. The
Karush-Kuhn-Tucker conditions \cite{KKT} are given by%
\begin{eqnarray}
\frac{\partial L}{\partial p_{1}} &=&2p_{1}\mathbf{r}_{1}^{T}\mathbf{r}_{1}+2p_{2}\mathbf{r}_{1}^{T}\mathbf{r%
}_{2}-2\mathbf{r}_{o}^{T}\mathbf{r}_{1}-\lambda _{1}+\lambda =0,  \notag \\
\frac{\partial L}{\partial p_{2}} &=&2p_{1}\mathbf{r}_{1}^{T}\mathbf{r%
}_{2}+2p_{2}\mathbf{r}_{2}^{T}\mathbf{r}_{2}-2\mathbf{r}_{o}^{T}\mathbf{r}_{2}-\lambda _{2}+\lambda =0,  \notag \\
\lambda _{i}p_{i} &=&0,\lambda _{i}\geq 0,p_{i}\geq 0,\sum p_{i}-1=0.
\label{lll2}
\end{eqnarray}
Solving above Eq. (\ref{lll2}) by $\frac{\partial L}{\partial p_{1}}-\frac{\partial L%
}{\partial p_{2}}=0$, we have%
\begin{eqnarray}
2(p_{1}\mathbf{r}_{1}+p_{2}\mathbf{r}_{2})^{T}(\mathbf{r}_{1}-\mathbf{r}_{2})&=&2\mathbf{r}_{o}^{T}(\mathbf{r}_{1}-\mathbf{r}_{2})+\lambda_{1}-\lambda_{2}   \notag \\
p_{1}+p_{2} &=&1.  \label{l3}
\end{eqnarray}%
Then it will arrive at the valid $p_{i}$ as%
\begin{eqnarray}
p_{1} &=&\frac{\left( \mathbf{r}_{o}-\mathbf{r}_{2} \right)^{T}\left( \mathbf{r}_{1}-\mathbf{r}_{2} \right) +\lambda_{1}-\lambda_{2}}{\Vert\mathbf{r}_{1}-\mathbf{r}_{2}\Vert_{2}^{2}}      \notag \\
p_{2} &=&1-\tilde{p}_{1}.  \label{l4}
\end{eqnarray}%
If $0\leq(\mathbf{r}_{o}-\mathbf{r}_{2})^{T}(\mathbf{r}_{1}-\mathbf{r}_{2})\leq\Vert\mathbf{r}_{1}-\mathbf{r}_{2}\Vert_{2}^{2}   $, we can get $\lambda_{1}=\lambda_{2}=0$.
Insert $\tilde{p}_{i}$ into $\chi _{1,2}\left( \vec{p}\right) $, one will
find%
\begin{eqnarray}
D^{2}(\rho,\chi_{1,2}\left( \vec{p}\right))=\Vert\mathbf{r}_{o}-\mathbf{r}_{2}\Vert_{2}^{2}-\frac{\left[  \left( \mathbf{r}_{o}-\mathbf{r}_{2} \right)^{T}\left( \mathbf{r}_{1}-\mathbf{r}_{2} \right) \right] ^{2} }{\Vert\mathbf{r}_{1}-\mathbf{r}_{2}\Vert_{2}^{2}}.
\end{eqnarray}%
If $(\mathbf{r}_{o}-\mathbf{r}_{2})^{T}(\mathbf{r}_{1}-\mathbf{r}_{2})<0$, we can get
\begin{eqnarray}
\lambda_{1}&=&\frac{-\left( \mathbf{r}_{o}-\mathbf{r}_{2} \right)^{T}\left( \mathbf{r}_{1}-\mathbf{r}_{2} \right)}{\Vert\mathbf{r}_{1}-\mathbf{r}_{2}\Vert_{2}^{2}}, \notag\\
p_{2}&=&1.
\end{eqnarray}%
The optimal distance is
\begin{eqnarray}
D(\rho ,\chi _{1,2}\left( \vec{p}\right) )=\Vert\mathbf{r}_{o}-\mathbf{r}_{2}\Vert_{2}.
\end{eqnarray}%
If $(\mathbf{r}_{o}-\mathbf{r}_{2})^{T}(\mathbf{r}_{1}-\mathbf{r}_{2})>\Vert\mathbf{r}_{1}-\mathbf{r}_{2}\Vert_{2}^{2}$, we can obtain
\begin{eqnarray}
p_{1}&=&1\notag\\
\lambda_{2}&=&\frac{\left( \mathbf{r}_{o}-\mathbf{r}_{2} \right)^{T}\left( \mathbf{r}_{1}-\mathbf{r}_{2} \right)}{\Vert\mathbf{r}_{1}-\mathbf{r}_{2}\Vert_{2}^{2}}-1 .
\end{eqnarray}%
The optimal distance is
\begin{eqnarray}
D(\rho ,\chi _{1,2}\left( \vec{p}\right) )=\Vert\mathbf{r}_{o}-\mathbf{r}_{1}\Vert_{2}.
\end{eqnarray}%
The proof is completed.\hfill $\blacksquare $

A special case of theorem 1 is that the set $S$ is made up of two
orthonormal pure states $\left\vert \varphi _{1}\right\rangle $ and $%
\left\vert \varphi _{2}\right\rangle $. In this case, one can find that $%
\mathbf{r}_{1}^{T}\mathbf{r}_{2}=-1$ and $\left\vert \mathbf{r}%
_{o}^{T}\left( \mathbf{r}_{1}-\mathbf{r}_{2}\right) \right\vert \leq 1-%
\mathbf{r}_{1}^{T}\mathbf{r}_{2}$. Substituting these parameters into
Theorem 1, one will directly obtain the optimal distance as
\begin{equation}
D^{2}(\rho ,\chi _{1,2}\left( \vec{p}\right) )=\mathbf{r}_{o}^{T}\mathbf{r}%
_{o}-\left( \mathbf{r}_{o}^{T}\mathbf{r}_{i}\right) ^{2},  \label{chuizhi}
\end{equation}%
with $i=1,2$ and the optimal weight as
\begin{equation}
p_{i}=\frac{1}{2}\left( 1+\mathbf{r}_{o}^{T}\mathbf{r}_{i}\right) .
\label{chuizhip}
\end{equation}%
In particular, $\mathbf{r}_{o}^{T}\mathbf{r}_{i}=r_{o\alpha }r_{i\alpha }$,
if both $\left\vert \varphi _{1}\right\rangle $ and $\left\vert \varphi
_{2}\right\rangle $ are the eigenstates of one Pauli matrix $\sigma _{\alpha
}$ with $\alpha =x,y,z$, because the eigenstates of one Pauli matrix $\sigma
_{\alpha }$ lead to that $\left\langle \varphi _{i}\right\vert \sigma
_{\alpha }\left\vert \varphi _{i}\right\rangle =\pm 1$ and the vanishing
average values on other Pauli matrices. Eq. (\ref{chuizhi}) actually
presents an alternative quantifier of quantum coherence of the objective
state $\rho $ with respect to the basis defined by the two orthogonal pure
states in $S$, which coincides with the result in \cite{CT}. Besides, it is obvious that a quite neat result has been
given for the case with only two states in $S$. However, the cases for more
than two states in the set $S$ are much more complicated than that with $N=2$%
, which will be addressed in what follows.

\textbf{Theorem 2}.- If there are three quantum states ($N=3$) in the set $S$,
one can construct a pseudo-state $\chi _{1,2,3}\left( \vec{p}\right) $ with
the pseudo-probabilities as
\begin{eqnarray}
\tilde{p}_{1} &=&\frac{1}{d}(\mathbf{r}_{1}-\mathbf{r}_{2})^{T}\left[  (\mathbf{r}_{o}-\mathbf{r}_{2})(\mathbf{r}_{2}-\mathbf{r}_{3})^{T} \right. \notag \\
&&-\left.(\mathbf{r}_{2}-\mathbf{r}_{3})(\mathbf{r}_{o}-\mathbf{r}_{2})^{T}\right] (\mathbf{r}_{2}-\mathbf{r}_{3}) ,\label{P1} \\
\tilde{p}_{2} &=&\frac{1}{d}(\mathbf{r}_{1}-\mathbf{r}_{2})^{T}\left[  (\mathbf{r}_{1}-\mathbf{r}_{3})(\mathbf{r}_{o}-\mathbf{r}_{1})^{T} \right. \notag \\
&&-\left.(\mathbf{r}_{o}-\mathbf{r}_{1})(\mathbf{r}_{1}-\mathbf{r}_{3})^{T}\right] (\mathbf{r}_{1}-\mathbf{r}_{3}) ,  \label{P2} \\
\tilde{p}_{3} &=&1-\tilde{p}_{1}-\tilde{p}_{2},  \label{P3}
\end{eqnarray}%
where
\begin{eqnarray}
d =\Vert\mathbf{r}_{1}-\mathbf{r}_{2}\Vert_{2}^{2}\Vert\mathbf{r}_{3}-\mathbf{r}_{2}\Vert_{2}^{2}-\left[(\mathbf{r}_{1}-\mathbf{r}_{2})^{T}(\mathbf{r}_{3}-\mathbf{r}_{2}) \right] ^{2}.  \label{dd} 
\end{eqnarray}
If $\tilde{p}_{i}\geqslant 0$ holds for $i=1,2,3$, $\chi _{1,2,3}\left( \vec{%
p}\right) $ will be our expected optimal state with the optimal weights $%
p_{i}=\tilde{p}_{i}$, and the optimal distance%
\begin{equation}
D^{2}\left( \rho ,\chi _{1,2,3}\left( \vec{p}\right) \right)
=\sum_{ij}^{3}p_{i}p_{j}\mathbf{r}_{i}^{T}\mathbf{r}_{j}-2p_{i}\mathbf{r}%
_{o}^{T}\mathbf{r}_{i}+\mathbf{r}_{o}^{T}\mathbf{r}_{o}.  \label{T23}
\end{equation}
Otherwise, the optimal state and the closest distance are given by%
\begin{equation}
D(\rho ,\chi _{1,2,3}\left( \vec{p}\right) )=\min_{i<j}D(\rho ,\chi
_{i,j}\left( \vec{p}\right) ),i,j=1,2,3,  \label{J22a}
\end{equation}%
which as well as the corresponding weights can be solved by Theorem 1.

\textbf{Proof. }For $N=3$, one can substitute $\chi _{1,2,3}\left( \vec{p}%
\right) $ into $D^{2}(\rho ,\chi _{1,2,3}\left( \vec{p}\right) )$ and
establish the corresponding Lagrangian function as%
\begin{eqnarray}
L(p_{i},\lambda ,\lambda _{i}) &=&\sum_{ij}^{3}p_{i}p_{j}\mathbf{r}_{i}^{T}%
\mathbf{r}_{j}-2p_{i}\mathbf{r}_{o}^{T}\mathbf{r}_{i}+\mathbf{r}_{o}^{T}%
\mathbf{r}_{o}  \notag \\
&&-\sum_{i=1}^{3}\lambda _{i}p_{i}+\lambda (\sum_{i=1}^{3}p_{i}-1),
\end{eqnarray}%
where $\lambda $ and $\lambda _{i}$ are the Lagrangian multipliers. The
Karush-Kuhn-Tucker conditions are given by%
\begin{eqnarray}
\frac{\partial L}{\partial p_{i}} &=&2\sum_{j=1}^{3}p_{j}\mathbf{r}_{i}^{T}%
\mathbf{r}_{j}-2\mathbf{r}_{o}^{T}\mathbf{r}_{i}+\lambda -\lambda _{i}=0,
\label{L21} \\
\lambda _{i} &\geq &0,p_{i}\geq 0,\lambda _{i}p_{i}=0,i=1,2,3.  \notag
\end{eqnarray}
Solving  Eq. (\ref{L21}) by $\frac{\partial L}{\partial p_{1}}-\frac{\partial
L}{\partial p_{3}}=0$ and $\frac{\partial L}{\partial p_{2}}-\frac{\partial L%
}{\partial p_{3}}=0$, we have%
\begin{eqnarray}
2\left( p_{1}\mathbf{r}_{1}+p_{2}\mathbf{r}_{2}+p_{3}\mathbf{r}%
_{3}-\mathbf{r}_{o}\right)^{T} \left( \mathbf{r}_{1}-\mathbf{r}_{3}\right) &=&\lambda_{1}-\lambda_{3} ,  \notag \\
2\left( p_{1}\mathbf{r}_{1}+p_{2}\mathbf{r}_{2}+p_{3}\mathbf{r}%
_{3}-\mathbf{r}_{o}\right)^{T} \left( \mathbf{r}_{2}-\mathbf{r}_{3}\right) &=&\lambda_{2}-\lambda_{3} ,  \notag \\
p_{1}+p_{2}+p_{3} &=&1.  \label{L22}
\end{eqnarray}%
We rewrite Eq. (\ref{L22}) in matrix form $A\mathbf{P}=\mathbf{B}$ with%
\begin{equation}
A=\left(
\begin{tabular}{lll}
$\mathbf{r}_{1}^{T}\left( \mathbf{r}_{1}-\mathbf{r}_{3}\right) $ & $\mathbf{r%
}_{2}^{T}\left( \mathbf{r}_{1}-\mathbf{r}_{3}\right) $ & $\mathbf{r}%
_{3}^{T}\left( \mathbf{r}_{1}-\mathbf{r}_{3}\right) $ \\
$\mathbf{r}_{1}^{T}\left( \mathbf{r}_{2}-\mathbf{r}_{3}\right) $ & $\mathbf{r%
}_{2}^{T}\left( \mathbf{r}_{2}-\mathbf{r}_{3}\right) $ & $\mathbf{r}%
_{3}^{T}\left( \mathbf{r}_{2}-\mathbf{r}_{3}\right) $ \\
$1$ & $1$ & $1$%
\end{tabular}%
\right) ,
\end{equation}%
\begin{eqnarray}
\mathbf{P} &=&\left(
\begin{tabular}{lll}
$\tilde{p}_{1}$ & $\tilde{p}_{2}$ & $\tilde{p}_{3}$%
\end{tabular}%
\right) ^{T},  \notag \\
\mathbf{B} &=&\left(
\begin{tabular}{lll}
$\mathbf{r}_{o}^{T}\left( \mathbf{r}_{1}-\mathbf{r}_{3}\right)+\frac{1}{2}(\lambda_{1}-\lambda_{3}) $ \\
$\mathbf{r}_{o}^{T}\left( \mathbf{r}_{2}-\mathbf{r}_{3}\right)+\frac{1}{2}(\lambda_{2}-\lambda_{3}) $ \\
$1$
\end{tabular}%
\right).
\end{eqnarray}%
The determinant of matrix $A$ is%
\begin{eqnarray}
d =\Vert\mathbf{r}_{1}-\mathbf{r}_{2}\Vert_{2}^{2}\Vert\mathbf{r}_{3}-\mathbf{r}_{2}\Vert_{2}^{2}-\left[(\mathbf{r}_{1}-\mathbf{r}_{2})^{T}(\mathbf{r}_{3}-\mathbf{r}_{2}) \right] ^{2}.  \label{L24}
\end{eqnarray}%

\textit{Case 1}: $d=0$. The determinant $d = 0$ is equivalent to
\begin{equation}
\mathbf{r}_{1}-\mathbf{r}_{2}=\Delta(\mathbf{r_{3}-\mathbf{r}_{2}})
\end{equation}
where $\Delta$ is an arbitrary constant. Thus only two independent equations in Eq. (\ref{L22}) are left. We can always let some $p_{i}=0$ in $\mathbf{P}$. So the left two nonvanishing $%
p_{i}$ just correspond to the case of Theorem
1, and then we can choose the smallest one as
\begin{equation}
\min_{i<j}D(\rho ,\chi _{i,j}\left( \vec{p}\right) ),i,j=1,2,3,  \label{TT2}
\end{equation}%
which can be exactly solved by Theorem 1.

\textit{Case 2}: $d\neq0$. By calculating $\mathbf{P}=A^{-1}\mathbf{B}$, we
can get
\begin{eqnarray}
p_{1} &=&\tilde{p}_{1}+\frac{\mathbf{b}^{T}}{2d} \left[  \lambda_{1}\mathbf{b}-\lambda_{2}(\mathbf{a}+\mathbf{b})+\lambda_{3}\mathbf{a}\right]  ,\notag \\
p_{2} &=&\tilde{p}_{2}+\frac{(\mathbf{a}+\mathbf{b})^{T}}{2d} \left[ - \lambda_{1}\mathbf{b}+\lambda_{2}(\mathbf{a}+\mathbf{b})-\lambda_{3}\mathbf{a}\right] ,\notag \\
p_3&=&\tilde{p}_{3}+\frac{\mathbf{a}^{T}}{2d} \left[  \lambda_{1}\mathbf{b}-\lambda_{2}(\mathbf{a}+\mathbf{b})+\lambda_{3}\mathbf{a}\right] .
\end{eqnarray}%
with
\begin{eqnarray}
\tilde{p}_{1} &=&\frac{1}{d}   \left[ \mathbf{a}^{T}(\mathbf{c}\mathbf{b}^{T}-\mathbf{b}\mathbf{c}^{T})\mathbf{b}\right] \notag \\
\tilde{p}_{3} &=&\frac{1}{d}  \left[  \mathbf{a}^{T}(\mathbf{c}\mathbf{a}^{T}-\mathbf{a}\mathbf{c}^{T})\mathbf{b}\right]  \notag \\
\tilde{p}_{2} &=&1-\tilde{p}_{1}-\tilde{p}_{3}.
\end{eqnarray}%
and
\begin{eqnarray}
\mathbf{a} = \mathbf{r}_{1}-\mathbf{r}_{2},\mathbf{b} = \mathbf{r}_{2}-\mathbf{r}_{3},\mathbf{c} = \mathbf{r}_{o}-\mathbf{r}_{2}.
\end{eqnarray}%
If all $\tilde{p}_{i}$ $\in \lbrack 0,1]$ for $i=1,2,3$, then the optimal weights $p_{i}=%
\tilde{p}_{i}$. With the optimal weights $p_{i}$, one can calculate $\chi
_{1,2,3}\left( \vec{p}\right) =\sum_{i=1}^{3}p_{i}\rho_{i}$, and the optimal
distance is hence given by $D(\rho ,\chi _{1,2,3}\left( \vec{p}\right) ).$
If not all $\tilde{p}_{i}$ $\in \lbrack 0,1]$, the optimal weights should be
on the boundaries which are described by (i) $p_{1}=0$, $\lambda _{1}>0$; (ii)$p_{2}=0$, $\lambda _{2}>0$;
(iii)$p_{3}=0$, $\lambda _{3}>0$. In other words, there
exists at least one $p_{i}=0$ among the three. This corresponds to the case of Theorem 1 again. The final optimal distance is shown in Eq. (\ref{TT2}). The proof is completed.\hfill $\blacksquare $

Theorem 2 shows some kind of degradation relation. That is, the validity of
the pseudo-probabilities have to be evaluated first. Once the
pseudo-probabilities are invalid, the problem with $N=3$ will be reduced
into the case with $N=2$, which implies that one has to use theorem 2 three
times to find the minimal value. Such a degradation relation ensures that
Theorem 2 for $N=2$ will be automatically reduced to Theorem 1. Next, we
will focus on the case of $N=4$. To do so, we will have to first address the
special case of full-rank $4\times K$ matrix $A$, which is defined as $%
A_{ij}=\mathrm{Tr}(\sigma_{i}\rho_{j})$ with $j=1,2,3,\cdots ,K$ and $i=1,2,3,4$ corresponding
to $x,y,z$, and $\sigma _{4}=\left(
\begin{array}{cc}
1 & 0 \\
0 & 1%
\end{array}%
\right) $.

\textbf{Theorem 3}.-If there are four quantum states ($N=4$) in the set $S$
with the rank-$4$ matrix $A$, one can obtain the exact objective state $\rho
$ with the pseudo-probabilities
\begin{equation}
\mathbf{\tilde{p}=}A^{-1}\mathbf{\tilde{r}}
\end{equation}%
where $\mathbf{\tilde{r}=[r};1]$ is a four-dimensional vector. If $\tilde{p}%
_{j}\in \lbrack 0,1]$ for all $j$, the objective state $\rho $ can be
exactly written as the convex sum of the four quantum states with the optimal
weights $\mathbf{p=\tilde{p}}$. Otherwise, the optimal approximation can be
solved based on Theorem 2 as
\begin{equation}
\min_{i<j<k}D(\rho ,\chi _{i,j,k}\left( \vec{p}\right) ),i,j,k=1,2,3,4.
\end{equation}

\textbf{Proof}. The convex optimization guarantees an important property
that if the optimal solutions of a problem with some constraints $S_{1}$
don't satisfy the other constraints $S_{2}$, then the optimal solutions with
the constraints $S_{1}\cup $ $S_{2}$ must be present on the boundaries of $%
S_{2}$. Now let's suppose that the objective quantum state $\rho $ can be
exactly expressed by the linear sum of the given four quantum states in $S$,
then one will definitely find the pseudo-probabilities $\tilde{p}_{i}$ such
that $\rho =\sum_{i=1}^{4}\tilde{p}_{i}\rho_{i}$ with $\sum\limits_{i}\tilde{p}_{i}=1$.
Note that $\tilde{p}_{i}$ could be negative. Thus in the Bloch
representation, the above relation can be rewritten as
\begin{equation}
A\mathbf{\tilde{p}=\tilde{r}}  \label{pseu}
\end{equation}%
with $\mathbf{\tilde{r}=[r};1\mathbf{]}$ and $\mathbf{\tilde{p}}=\left(
\tilde{p}_{1},\tilde{p}_{2},\tilde{p}_{3},\tilde{p}_{4}\right) ^{T}$ . Since
$A$ is of full rank, Eq. (\ref{pseu}) has a unique solution as $\mathbf{%
\tilde{p}=}A^{-1}\mathbf{\tilde{r}}$. It is obvious that if $\tilde{p}%
_{i}\in \lbrack 0,1]$ for all $i$, one will directly obtain the optimal
solution  $\mathbf{p=\tilde{p}=}A^{-1}\mathbf{%
\tilde{r}}$. However, if there exist $\tilde{p}_{i}\notin \lbrack 0,1]$,
this means the optimal solution doesn't satisfy the constraints $p_{i}\in
\lbrack 0,1]$. So the optimal solutions can only be present at the
boundaries of $p_{i}\in \lbrack 0,1]$. Following the Lagrangian multiplier
method similar to the proof of Theorem 2, the Karush-Kuhn-Tucker condition
\cite{KKT} indicates that the optimal distance should be
\begin{equation}
\min_{i<j<k}D(\rho ,\chi _{i,j,k}\left( \vec{p}\right) ),i,j,k=1,2,3,4.
\end{equation}%
It is shown that the current optimal scheme for $N=4$ can be solved by
Theorem 2. The proof is completed. \hfill $\blacksquare $

\bigskip Up to now, we have entirely solved our optimal scheme with $N\leq 3$
and $N=4$ with the full-rank matrix $A$. To proceed, we will have to first
introduce a quite vital theorem 4, which is another cornerstone of our paper.

\textbf{Theorem 4}.- Suppose $\left\{ p_{i},\rho_{i}\right\rbrace_{N} $ to be the quantum state decomposition of a
qubit state $\rho =\sum_{i=1}^{N}p_{i}\rho_{i}$ with $N\geq 4$. Let $R$ denote the rank
of the $4\times N$ matrix $A$ (obviously $R\leqslant 4$), then $R$ states as
$\left\{ q_{j},\rho_{j}\right\} _{R}$ can always
be found from $\left\{ p_{i},\rho_{i}\right\rbrace_{N} $
to make another quantum state decomposition such that $\rho
=\sum_{j=1}^{R}q_{j}\rho_{j} $ where $j=1,2,\cdots ,R$ implies reordering the $R$ states
and $R=1$ means a quantum state as $\rho =\rho_{i}$.

\textbf{Proof. }Since the matrix $A$ is of rank $R$, one can always find $R$
linearly independent columns from $A$  and one additional column of $A$ to construct a $4\times
(R+1)$ matrix denoted by $\tilde{A}$. Here we'd like to use $\tilde{A}_i$ to represent the $i$th column of $\tilde{A}$ with $i=1,2,\cdots, R+1$. Additionally, 
we define $\tilde{r}_o=[\mathbf{r}_o;1]$ for convenience. It is obvious that the $R+1$ columns in $\tilde{A}$ are linearly dependent, namely, there exist $k_i, i=1,2,\cdots, R+1$ such
 that $\sum_{i=1}^{R+1} k_{i}\tilde{A}_{i} =0$, which implies  $\sum_{i=1}^{R+1}
k_{i}=0$ due to $A_{i4}=1$. In addition, Caratheodory theorem \cite{Ca1,Ca2} shows 
that  $\tilde{r} _o$ can be written as the convex mixing of the $R+1$ columns of $\tilde{A}$, i.e., $\tilde{r} _o=\sum_{i=1}^{R+1}
p_{i}\tilde{A}_{i} $ with $\sum_ip_i=1$ and $ \tilde{p}_i\geq 0$.Thus one can obtain
\begin{equation}
\tilde{r} _o=\sum_{i=1}^{R+1}p_{i}\tilde{A}_{i} -\alpha \sum_{i=1}^{R+1} k_{i}\tilde{A}_{i}  
=\sum_{i=1}^{R+1}p_{i}(1-\alpha\frac{k_i}{p_i})\tilde{A}_{i}. \label{top}
\end{equation}%
Let $\alpha=\frac{p_{i'}}{\tilde{k}_{i'}}=\min_{1\leq i \leq R+1}\left\lbrace \frac{p_i}{\tilde{k}_i}|k_{i}> 0 \right\rbrace$, we will find that $1-\alpha\frac{k_{i'}}{p_{i'}}\left \{
\begin{array}{cc}
  =0&   i=i'  \\
 \geq 0 &i\neq i' 
    \end{array}
\right.$ and $\sum_i p_i(1-\alpha\frac{k_{i}}{p_{i}})=1$ which mean that at most $R$ columns of $\tilde{A}$ are enough to convexly construct $\tilde{r} _o$. 
The proof is finished. 				\hfill $\blacksquare $

The importance of Theorem 4 is twofold. One is that if a density matrix can
be written as the convex mixing of $N\geq 4$ quantum states, Theorem 4 shows
that $R\leqslant 4$ quantum states are enough to prepare the given density
matrix. The other is that the $R\leqslant 4$ quantum states aren't so arbitrary
that they can be exactly found from the previous $N\geq 4$ quantum states. In
other words, if $N\geq 4$ quantum states in set $S$ can optimally construct a
given density matrix, one can select $R\leqslant 4$ from the $N$ quantum states
to construct the given density matrix with an equal optimization degree.
This can also be given in the following rigorous way.

\textbf{Theorem 5}.- For $N\geqslant 4$ with $4\times N$\ rank-$R$ matrix $A$%
, \ the optimal approximation is determined by%
\begin{equation}
\min_{i_{1}<i_{2}<\cdots <i_{R}}D(\rho ,\chi _{i_{1},i_{2},\cdots
,i_{R}}\left( \vec{p}\right) ),i_{\alpha}=1,2,\cdots,N.
\end{equation}

\textbf{Proof}. The proof of this theorem is straightforward. Suppose we can select $%
N^{\prime }\geqslant R$ quantum states from the set $S$ to prepare the state
denoted by $\rho _{o}=\sum\limits_{i=1}^{N^{\prime }}p_{i}^{\prime
}\rho_{i} $ such
that $D(\rho ,\rho _{o})$ achieves the optimal distance. Theorem 4 indicates
that one can always find $R$ states from the selected $N^{\prime }$ quantum
states to exactly prepare the state $\rho _{o}$ with some proper weights.
This implies that for $N\geqslant 4$, the convex mixing of only $R$ states
in $S$ is enough to achieve the optimal distance. Therefore, we can directly
consider all potential combinations of only $R$ quantum states among the set $S$%
. The minimal distance will give our expected optimal result, which is
exactly what Theorem 5 says.  \hfill $\blacksquare $

\textit{Examples.-} To further demonstrate the validity of our theorems and
their applications, we will consider various randomly generated objective
states and set $S$ by comparing our analytical results with the numerical
ones. For simplicity, we'd like to write the objective density matrix as $%
\rho =\left(
\begin{array}{cc}
1-a & k\sqrt{a(1-a)}e^{-i\Phi } \\
k\sqrt{a(1-a)}e^{i\Phi } & a%
\end{array}%
\right) $ with $k,a\in \lbrack 0,1]$ and $\Phi $ denoting the phase, and the
explicit forms of $S=\left\{ \rho_{1},\rho_2 ,\cdots ,\rho_N\right\} $ in all the below examples are given in the
Appendix A.
\begin{figure}[tbp]
\centering
\includegraphics[width=0.48\columnwidth,height=1.3in]{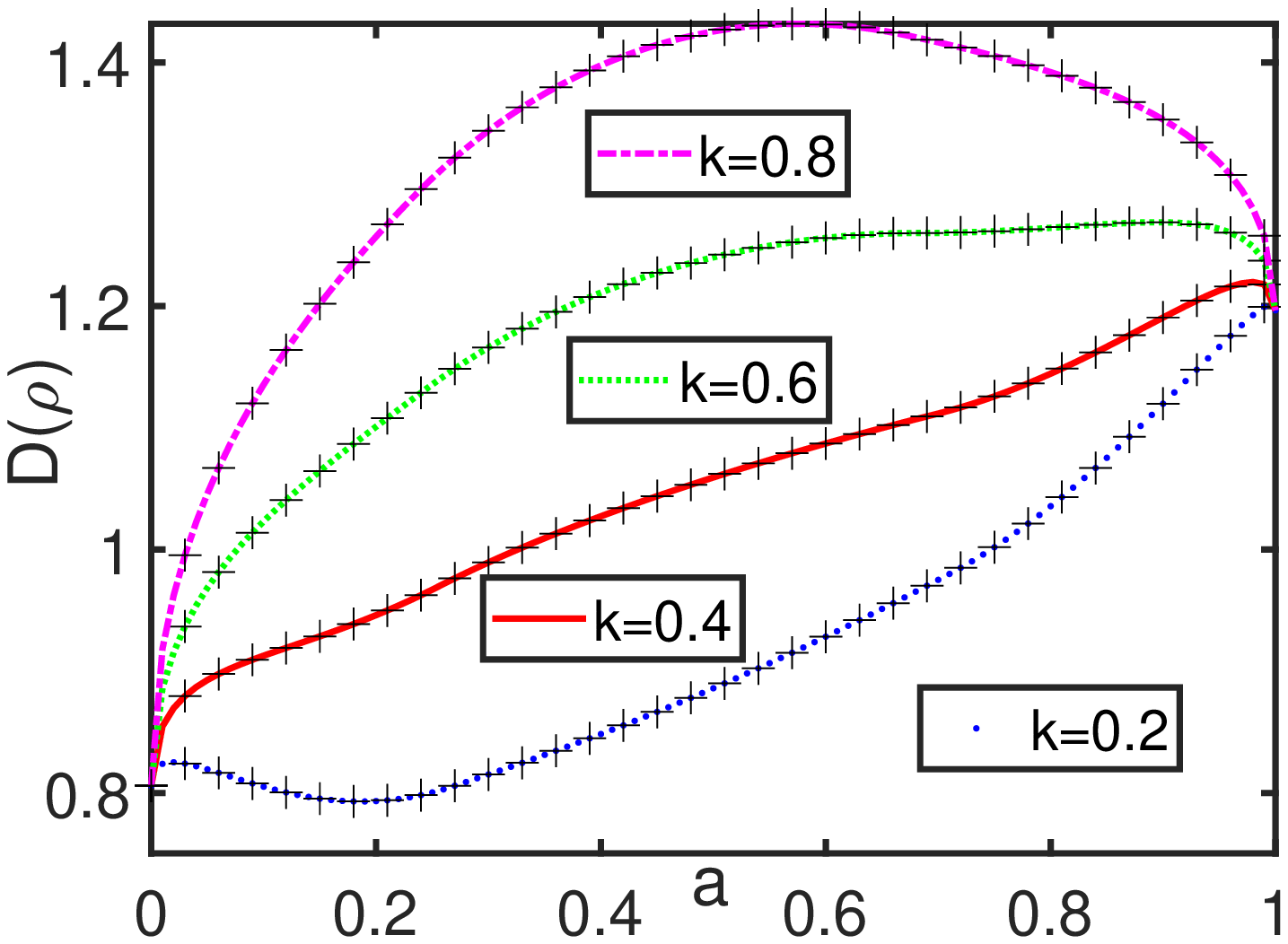} %
\includegraphics[width=0.48\columnwidth,height=1.3in]{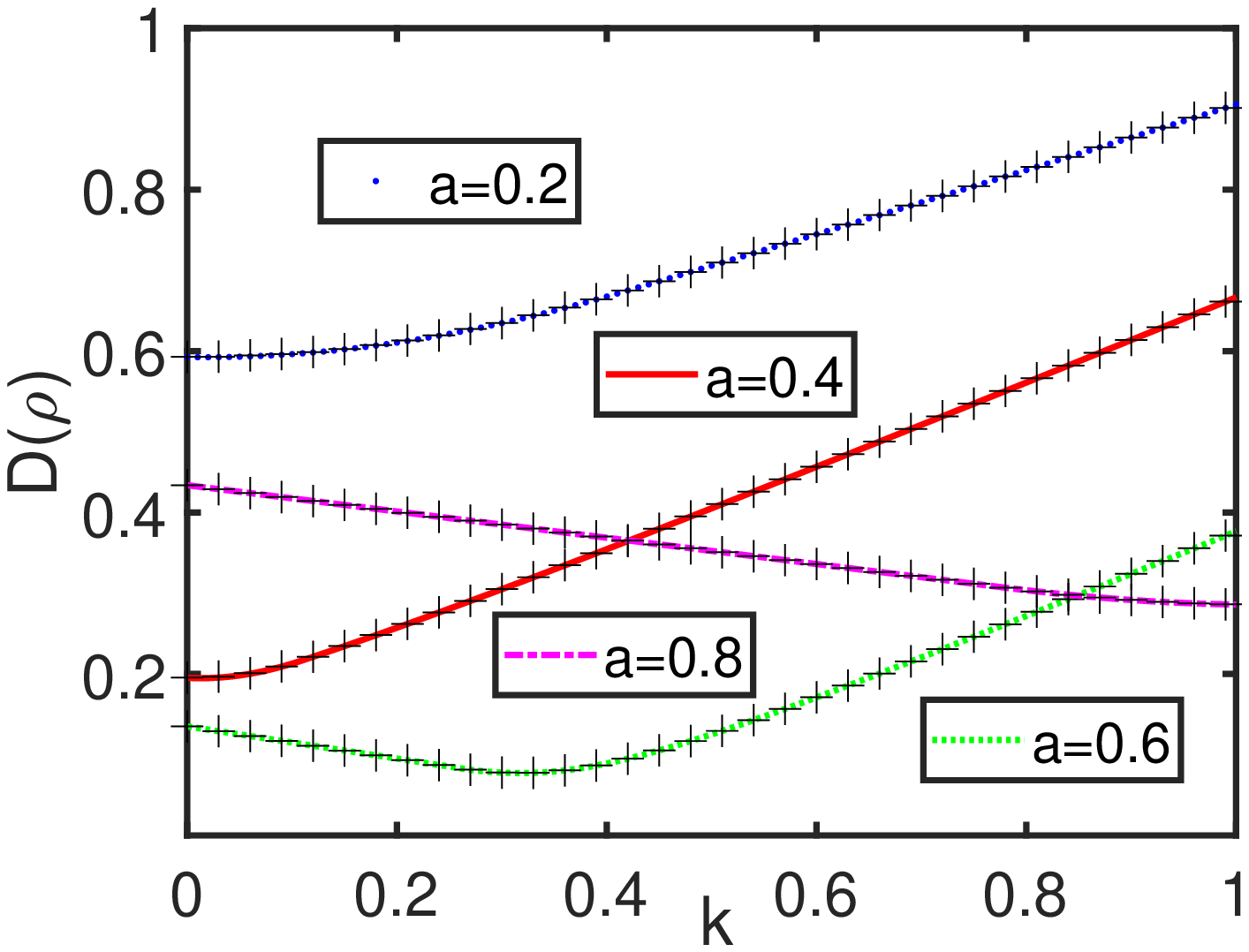} %
\includegraphics[width=0.48\columnwidth,height=1.3in]{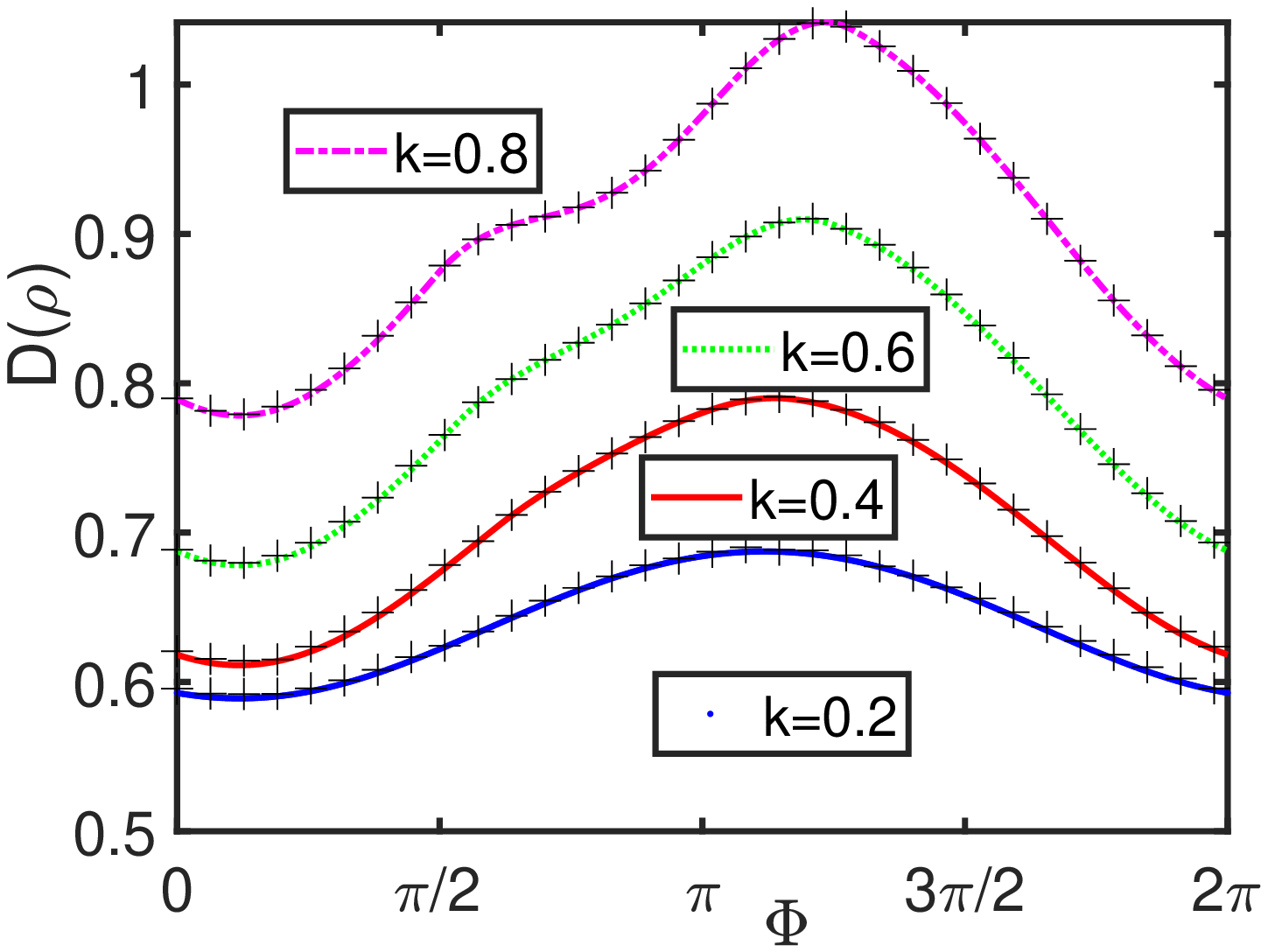} %
\includegraphics[width=0.48\columnwidth,height=1.3in]{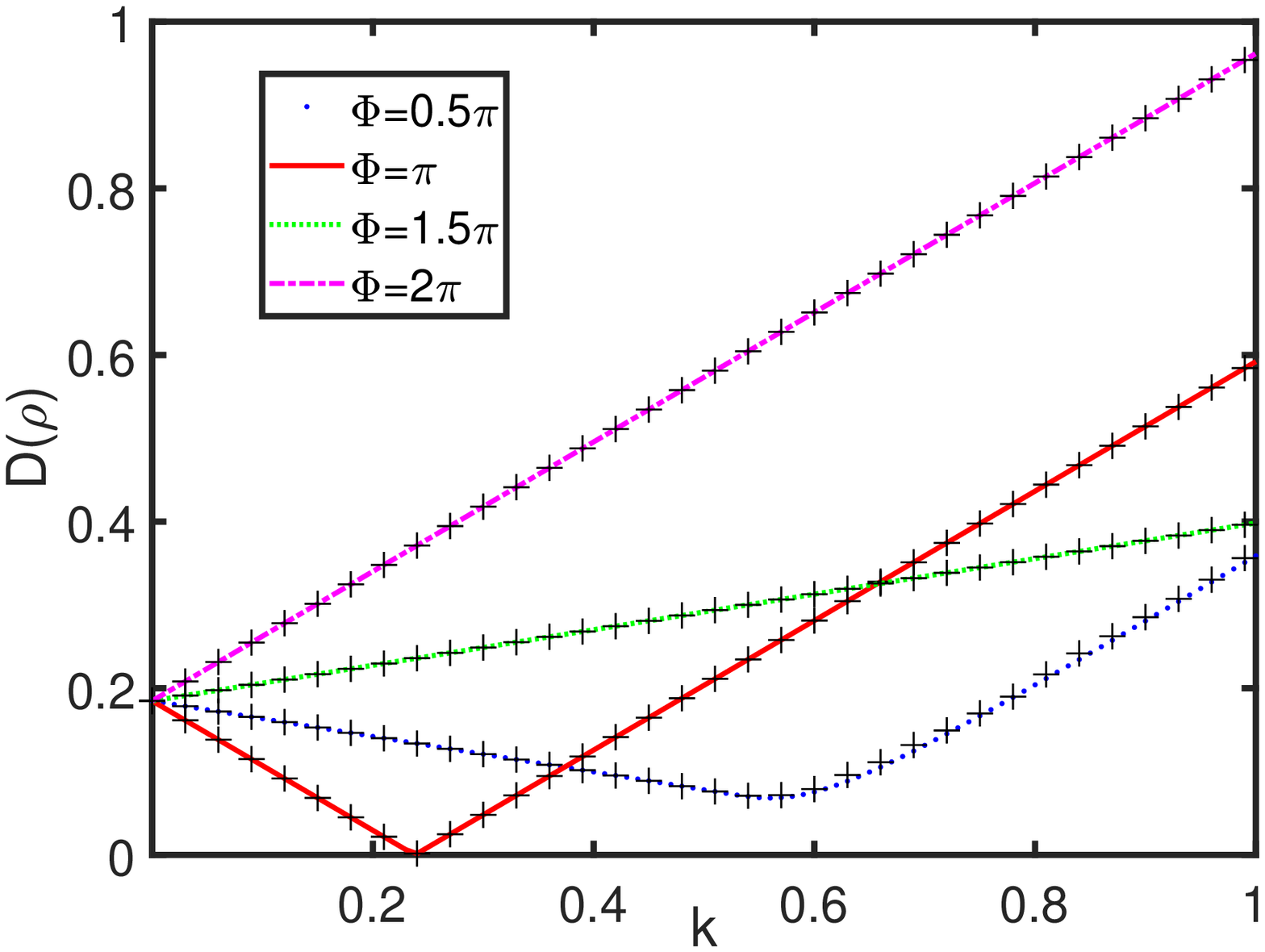}
\caption{(color online) The optimal distance $D(\protect\rho )$ versus
various parameters $a$ and $k$. The solid line corresponds to the strictly
analytical expressions given in our theorems, while the numerical solutions
are marked with "+".}
\end{figure}

(i) $N=2$. The random phase for the objective state $\rho $ is given by $%
\Phi =1.3580\pi $ with $\{k=0.2,0.4,0.6,0.8\}$. The optimal distance denoted
by $D(\rho )$ versus $a\in \lbrack 0,1]$ is plotted in Fig. 1 (a), which
shows the perfect consistency between the numerical and the analytical
results, and further supports our theorem 1.

(ii) $N=3$. The objective state $\rho $ is generated with a random phase $%
\Phi =0.4511\pi $ and $\{a=0.2,0.4,0.6,0.8\}$. The optimal distance $D(\rho
) $ versus $k\in \lbrack 0,1]$ is plotted in Fig. 1 (b), which validates our
theorem 2 based on the perfect consistency.

(iii) $N=4$. The objective state $\rho $ is generated by the randomly
generated $a=0.7522$ with $\{k=0.2,0.4,0.6,0.8\}$. The optimal distance $D(\rho )$ versus $\Phi\in \lbrack 0,2\pi]$ is plotted
in Fig. 1 (c). This indicates the perfect consistency with our theorem 3.

(iv) $N=4$ with two pairs of orthonormal states. The objective state $\rho $
is given by a random $a=0.3135$ and $\{\Phi =\frac{1}{2}\pi ,\pi ,\frac{3}{2}%
\pi ,2\pi \}$. We set $\left\langle \varphi _{1}\right\vert \left. \varphi
_{2}\right\rangle =0$ and $\left\langle \varphi _{3}\right\vert \left.
\varphi _{4}\right\rangle =0$ which can be found in the Appendix B. The
optimal distance $D(\rho )$ versus $k\in \lbrack 0,1]$ is plotted in Fig. 1
(d), which is perfectly consistent with our theorem 4.

(v) $N=5$. The objective state $\rho $ is established with $k=0.5625$, $%
\{a=0.2,0.4,0.6,0.8\}$ and the free $\Phi \in \lbrack 0,2\pi ]$. The optimal
distance $D(\rho )$ versus $\Phi$ is plotted in Fig. 2 (a), which shows our
theorem 4 and theorem 5 are valid.

(vi) $N=10$. To further illustrate the application of our main results to
large-scale cases. We consider the quantum state set $S$, which includes ten
states. The objective state $\rho $ is determined by the random$\{\Phi =\frac{1}{2}\pi ,\pi ,\frac{3}{2}%
\pi ,2\pi \}$ and the free $a \in \lbrack 0,1 ]$. The optimal
distance $D(\rho )$ versus $a $ is plotted in Fig. 2 (b).
\begin{figure}[tbp]
\centering
\includegraphics[width=0.48\columnwidth,height=1.3in]{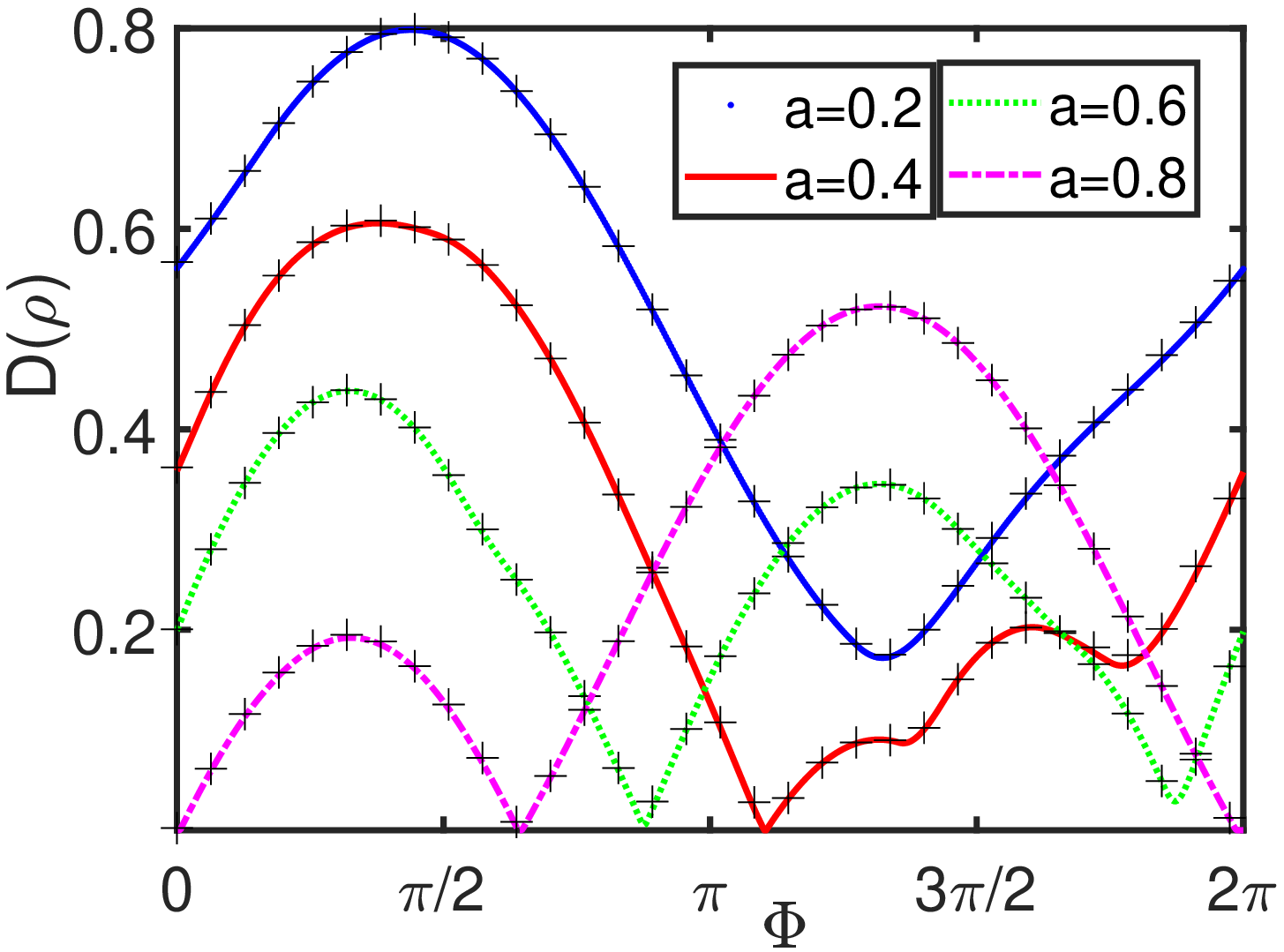} %
\includegraphics[width=0.48\columnwidth,height=1.3in]{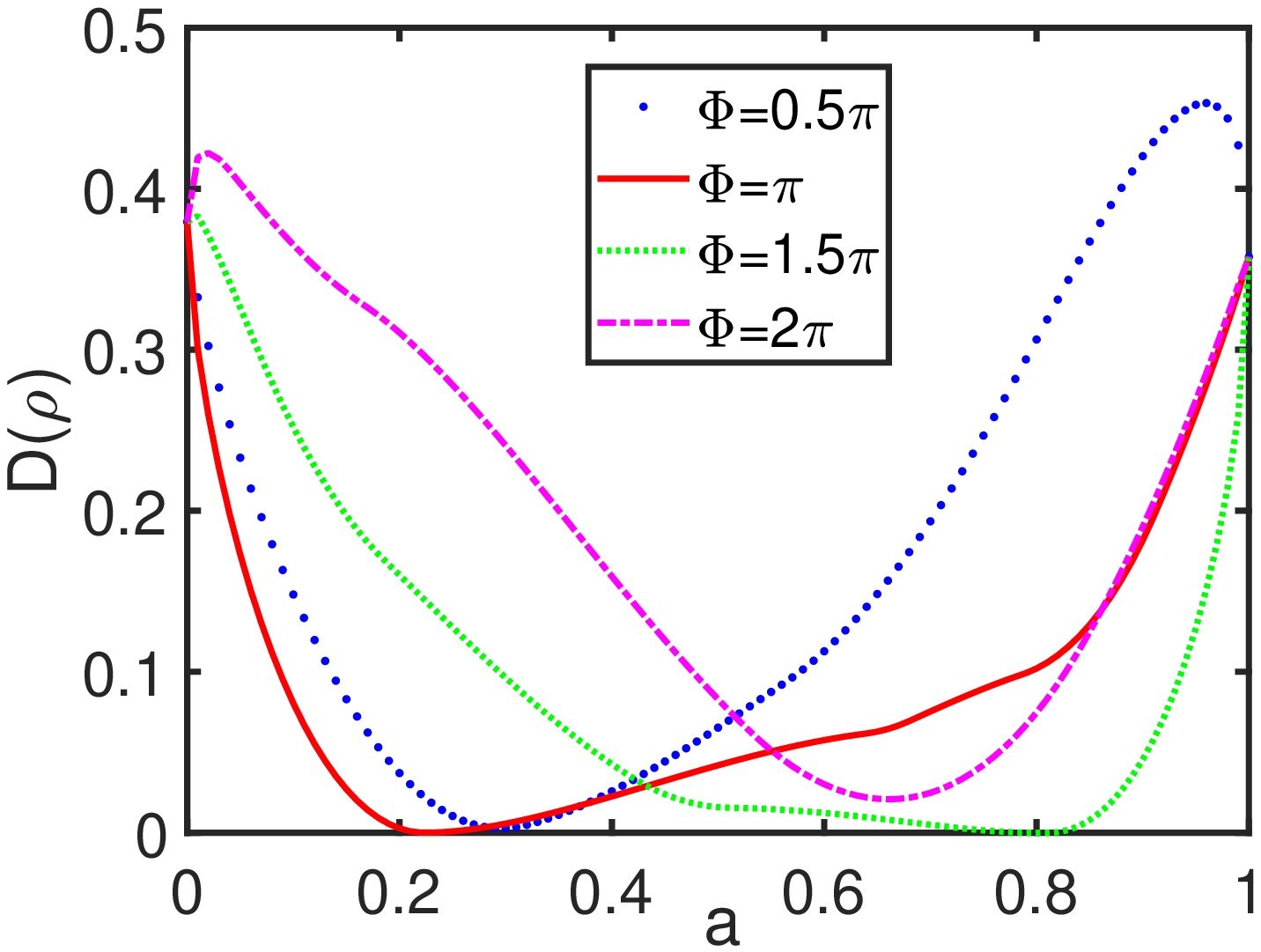}
\caption{(color online) The optimal distance $D(\protect\rho )$ versus $\Phi
$. The strictly analytical solutions correspond to the solid line and the
numerical solutions are marked with "+".}
\end{figure}

(vii) \textit{The set $S$ of the eigenstates of Pauli matrices}.- Besides
the above numerical tests, we give an analytic example. The pure-state set $%
S $ is made up of the eigenstates of two Pauli matrices $\sigma _{x}$ and $%
\sigma _{y}$, i.e., $\left\vert \varphi _{1\pm }\right\rangle =\left( 1,\pm
1\right) ^{T}/\sqrt{2}$, $\left\vert \varphi _{2\pm }\right\rangle =\left(
1,\pm i\right) ^{T}/\sqrt{2}$. In this case, one can find that the rank of
matrix $A$ is equal to 3, so based on our theorems, we only require three
among the four quantum states to solve the optimal scheme. Since the
Pauli-distance is invariant under the state transformations $\rho (a,k,\Phi
)\rightarrow \rho (a,k,\Phi \pm n\pi /2)$ (with the integer $n$), it is
enough only to consider $\Phi \in \left[ 0,\pi /2\right] $. If $\Lambda
=r_{ox}+r_{oy}\leq 1$, one can calculate the optimal distance as%
\begin{equation}
D(\rho ,\chi _{1,2,3,4}\left( \vec{p}\right) )=r_{oz},
\end{equation}%
and the optimal weights as%
\begin{eqnarray}
p_{1+} &=&r_{ox},p_{1-}=0,  \notag \\
p_{2-} &=&\frac{1}{2}\left[ 1-\Lambda \right] ,p_{1+}=1-p_{1+}-p_{2-},
\end{eqnarray}%
or%
\begin{eqnarray}
p_{2+} &=&r_{oy},p_{2-}=0, \\
p_{1-} &=&\frac{1}{2}\left[ 1-\Lambda \right] ,p_{1+}=1-p_{1-}-p_{2+};
\notag
\end{eqnarray}%
If $\Lambda >1$, the optimal distance is given by%
\begin{equation}
D^{2}(\rho ,\chi _{1,2,3,4}\left( \vec{p}\right) )=r_{oz}^{2}+\frac{1}{2}%
\left( r_{ox}+r_{oy}-1\right) ^{2},
\end{equation}%
and the optimal weights as%
\begin{eqnarray}
p_{1+} &=&\frac{1}{2}\left[ 1+r_{ox}-r_{oz}\right] , \\
p_{2+} &=&1-p_{1+},p_{1-}=p_{2-}=0.  \notag
\end{eqnarray}%
Notice the identities%
\begin{eqnarray}
r_{ox} &=&2k\sqrt{a\left( 1-a\right) }\cos \Phi ,  \notag \\
r_{oy} &=&2k\sqrt{a\left( 1-a\right) }\sin \Phi ,  \notag \\
r_{oz} &=&1-2a.
\end{eqnarray}%
Similar results can be found in \cite{CC2}.
The whole story about the various cases of the Pauli-distance is
given in the Appendix B.

\emph{Conclusions and discussion}.- To sum up, we have thoroughly solved the optimal approximation of a qubit state with the given quantum state set. The key points are that this type of
problems is a convex optimization problem, and the optimal approximation
with $N\geq 4$ quantum states can be converted into the problem with not more
than four quantum states. Thus we successfully solve the cases with not more
than four quantum states and equivalently solve the whole question.
We'd like to mention that an interesting link between the optimal approximation and the triple uncertainty relation was given in Ref. \cite{FSM} , where the state set with four pure states was also addressed. However, the four (or six) qubit pure states considered in Ref. \cite{FSM} are two (or three) pairs of orthogonal qubit pure states restricted in real space. Our current work extends the given state set to include arbitrary number of qubit states and especially loosens the restriction not only from real to complex quantum states, but also from pure to mixed states. 

Finally, we emphasize that the distance here is based on the
trace norm of two states. Other distances of states should also be feasible
and be worth studying. Besides, we only consider the optimal approximation
of a qubit density matrix. An intuitive observation shows that it is hard to
obtain the analytic solution in high-dimensional systems, however, there are
many particular and interesting cases requiring further research, including
a simple proof for the positive Hessian matrix. For example, our corollary
can be understood as a coherence measure of a qubit state (our
theorem 2 gives the more general case similar to \cite{C13}). It would be
interesting to approximate an arbitrary quantum state by the orthonormal
pure-state set. It could be more significant if we consider the
approximation of a density matrix of composite systems by local or nonlocal
state set.

\acknowledgments This work was supported by the National Natural Science
Foundation of China under Grant No.11775040, No. 12011530014 and
No.11375036, and the Fundamental Research Fund for the Central Universities
under Grant No. DUT20LAB203.

\appendix

\section{ The explicit forms of the set $S$ for examples}

\textit{In example (i)}, the quantum state set $S$ includes two randomly
generated quantum states whose Bloch vectors are
\begin{eqnarray}
\mathbf{r}_{1} &=&\left(
\begin{tabular}{lll}
$0.7888$ & $0.1788$ & $-0.1182$%
\end{tabular}%
\right) ^{T},  \notag \\
\mathbf{r}_{2} &=&\left(
\begin{tabular}{lll}
$0.4715$ & $ 0.4288$ & $0.5066$%
\end{tabular}%
\right) ^{T}. 
\end{eqnarray}%

\textit{In example (ii)}, the quantum state set $S$ is given in Bloch representation as
\begin{eqnarray}
\mathbf{r}_{1} &=&\left(
\begin{tabular}{lll}
$-0.0347$ & $0.0178$ & $0.0088$%
\end{tabular}%
\right) ^{T},  \notag \\
\mathbf{r}_{2} &=&\left(
\begin{tabular}{lll}
$0.3369$ & $-0.7514$ & $-0.2106$%
\end{tabular}%
\right) ^{T},  \notag \\
\mathbf{r}_{3} &=&\left(
\begin{tabular}{lll}
$0.3784$ & $0.8012$ & $-0.4636$%
\end{tabular}%
\right) ^{T}. 
\end{eqnarray}%
The quantum state $\mathbf{r}_{3}$ is a pure state.

\textit{In example (iii)}, the four randomly generated quantum states in the
set $S$ are with
\begin{eqnarray}
\mathbf{r}_{1} &=&\left(
\begin{tabular}{lll}
$-0.3384$ & $0.2543$ & $0.4151$%
\end{tabular}%
\right) ^{T},  \notag \\
\mathbf{r}_{2} &=&\left(
\begin{tabular}{lll}
$0.6385$ & $0.0199$ & $0.5333$%
\end{tabular}%
\right) ^{T},  \notag \\
\mathbf{r}_{3} &=&\left(
\begin{tabular}{lll}
$0.1693$ & $0.0661$ & $0.0845$%
\end{tabular}%
\right) ^{T},  \notag \\
\mathbf{r}_{4} &=&\left(
\begin{tabular}{lll}
$0.5275$ & $-0.2657$ & $0.5443$%
\end{tabular}%
\right) ^{T}. 
\end{eqnarray}%

\textit{In example (iv)}, the two pairs of randomly generated orthonormal
pure states in $S$ are given by
\begin{eqnarray}
\left\vert \varphi _{1}\right\rangle &=&\left(
\begin{array}{c}
0.9531-0.1605i \\
-0.2363+0.1001i%
\end{array}%
\right) ,  \notag \\
\left\vert \varphi _{2}\right\rangle &=&\left(
\begin{array}{c}
0.2495+0.0605i \\
0.9665+0.0028i%
\end{array}%
\right) ,  \notag \\
\left\vert \varphi _{3}\right\rangle &=&\left(
\begin{array}{c}
-0.6140+0.1652i \\
-0.4624-0.6179i%
\end{array}%
\right) ,  \notag \\
\left\vert \varphi _{4}\right\rangle &=&\left(
\begin{array}{c}
-0.3277+0.6988i \\
0.6347+0.0374i%
\end{array}%
\right) .
\end{eqnarray}%
Note that $\left\langle \varphi _{1}\right\vert \left. \varphi
_{2}\right\rangle =0$ and $\left\langle \varphi _{3}\right\vert \left.
\varphi _{4}\right\rangle =0$.

\textit{In example (v)}, the quantum state set $S$ includes five randomly
generated quantum states which are
\begin{eqnarray}
\mathbf{r}_{1} &=&\left(
\begin{tabular}{lll}
$-0.4767$ & $0.5882$ & $-0.6051$%
\end{tabular}%
\right) ^{T},  \notag \\
\mathbf{r}_{2} &=&\left(
\begin{tabular}{lll}
$0.3041$ & $-0.2655$ & $0.1277$%
\end{tabular}%
\right) ^{T},  \notag \\
\mathbf{r}_{3} &=&\left(
\begin{tabular}{lll}
$0.0459$ & $-0.3519$ & $0.0202$%
\end{tabular}%
\right) ^{T},  \notag \\
\mathbf{r}_{4} &=&\left(
\begin{tabular}{lll}
$0.7263$ & $-0.1260$ & $-0.6758$%
\end{tabular}%
\right) ^{T},  \notag \\
\mathbf{r}_{5} &=&\left(
\begin{tabular}{lll}
$-0.5631$ & $-0.5566$ & $0.6108$%
\end{tabular}%
\right) ^{T}. 
\end{eqnarray}%
The quantum states $\mathbf{r}_{4}$ and $\mathbf{r}_{5}$ are pure states.

\textit{In example (vi)}, we consider the set $S$ composed of 10 randomly
generated quantum states which are listed as follows.
\begin{eqnarray}
\mathbf{r}_{1} &=&\left(
\begin{tabular}{lll}
$-0.0192$ & $ 0.5339$ & $0.4067$%
\end{tabular}%
\right) ^{T},  \notag \\
\mathbf{r}_{2} &=&\left(
\begin{tabular}{lll}
$-0.0299$ & $0.0694$ & $0.1474$%
\end{tabular}%
\right) ^{T},  \notag \\
\mathbf{r}_{3} &=&\left(
\begin{tabular}{lll}
$0.1865$ & $-0.2202$ & $-0.0863$%
\end{tabular}%
\right) ^{T},  \notag \\
\mathbf{r}_{4} &=&\left(
\begin{tabular}{lll}
$0.4860$ & $-0.3405$ & $-0.5005$%
\end{tabular}%
\right) ^{T},  \notag \\
\mathbf{r}_{5} &=&\left(
\begin{tabular}{lll}
$-0.4864$ & $-0.4754$ & $-0.4707$%
\end{tabular}%
\right) ^{T}, 
\end{eqnarray}%
and
\begin{eqnarray}
\mathbf{r}_{6} &=&\left(
\begin{tabular}{lll}
$-0.5738$ & $-0.0250$ & $0.5583$%
\end{tabular}%
\right) ^{T},  \notag \\
\mathbf{r}_{7} &=&\left(
\begin{tabular}{lll}
$-0.0071$ & $0.0058$ & $-0.0298$%
\end{tabular}%
\right) ^{T},  \notag \\
\mathbf{r}_{8} &=&\left(
\begin{tabular}{lll}
$-0.5357$ & $-0.7338$ & $0.4177$%
\end{tabular}%
\right) ^{T},  \notag \\
\mathbf{r}_{9} &=&\left(
\begin{tabular}{lll}
$-0.9142$ & $-0.8936$ & $-0.8847$%
\end{tabular}%
\right) ^{T},  \notag \\
\mathbf{r}_{10} &=&\left(
\begin{tabular}{lll}
$0.4888$ & $0.8306$ & $0.2670$%
\end{tabular}%
\right) ^{T}. 
\end{eqnarray}%
The quantum states $\mathbf{r}_{8}$, $\mathbf{r}_{9}$ and $\mathbf{r}_{10}$ are pure states.

\section{Explicit calculation of the cases of Pauli matrices}

Let $\left\vert z\right\rangle _{s}=\left\{ \left\vert 0\right\rangle
,\left\vert 1\right\rangle \right\} $,$\ \left\vert x\right\rangle
_{s}=\left\{ \frac{1}{\sqrt{2}}(\left\vert 0\right\rangle \pm \left\vert
1\right\rangle )\right\} $ and $\left\vert y\right\rangle _{s}=\left\{ \frac{%
1}{\sqrt{2}}(\left\vert 0\right\rangle \pm i\left\vert 1\right\rangle
)\right\} $ be the eigenstates of the Pauli matrices $\sigma _{z}$, $\sigma
_{x}$ and $\sigma _{y}$, respectively. $s=\pm 1$ are the
eigenvalues. Since the Pauli-distance is invariant under the state
transformations $\rho (a,k,\Phi )\rightarrow \rho (1-a,k,\Phi )$ and $\rho
(a,k,\Phi )\rightarrow \rho (a,k,\Phi \pm n\pi /2)$ (with the integer $n$),
it is enough only to consider $a\in \left[ 0,1/2\right] $\ and $\Phi \in %
\left[ 0,\pi /2\right] $. We have%
\begin{eqnarray}
r_{ox} &=&2k\sqrt{a\left( 1-a\right) }\cos \Phi ,  \notag \\
r_{oy} &=&2k\sqrt{a\left( 1-a\right) }\sin \Phi ,  \notag \\
r_{oz} &=&1-2a.
\end{eqnarray}%
We study our scheme step by step from $N=3$, $N=4$ and $N=6$.

For $N=3$, we choose three of the six quantum states to solve the convex
optimization problem. This problem can be divided into the following four
cases.

\textbf{Case 1}.- The three selected quantum states are $\left\vert \varphi
_{1}\right\rangle =\left\vert \alpha \right\rangle _{1}$, $\left\vert
\varphi _{2}\right\rangle =\left\vert \alpha \right\rangle _{-1}$ and $%
\left\vert \varphi _{3}\right\rangle =\left\vert \alpha ^{\prime
}\right\rangle _{s_{3}}$ with $\left\{ \alpha \neq \alpha ^{\prime }\neq
\alpha ^{\prime \prime }|\alpha ,\alpha ^{\prime },\alpha ^{\prime \prime
}=x,y,z\right\} $. If $s_{3}=1$\ and$\ r_{o\alpha }+r_{o\alpha ^{\prime
}}\leq 1$, one can calculate the optimal distance as%
\begin{equation}
D(\rho ,\chi _{1,2,3}\left( \vec{p}\right) )=r_{o\alpha ^{\prime \prime }},
\label{c21}
\end{equation}%
and the optimal weights as
\begin{eqnarray}
p_{1} &=&1-p_{2}-p_{3},  \notag \\
p_{2} &=&\frac{1}{2}\left( 1-r_{o\alpha }-r_{o\alpha ^{\prime }}\right) ,
\notag \\
p_{3} &=&r_{o\alpha ^{\prime }}.  \label{c22}
\end{eqnarray}%
If $s_{3}=1$\ and $r_{o\alpha }+r_{o\alpha ^{\prime }}>1$, the optimal
distance is given by%
\begin{equation}
D^{2}(\rho ,\chi _{1,3}\left( \vec{p}\right) )=r_{o\alpha ^{\prime \prime
}}^{2}+\frac{1}{2}\left( r_{o\alpha }+r_{o\alpha ^{\prime }}-1\right) ^{2},
\label{c23}
\end{equation}%
and the optimal weights are given by%
\begin{eqnarray}
p_{1} &=&\frac{1}{2}\left( 1+r_{o\alpha }-r_{o\alpha ^{\prime }}\right) ,
\notag \\
p_{3} &=&1-p_{1},p_{2}=0,  \label{c24}
\end{eqnarray}%
If $s_{3}=-1$, the optimal distance is given by%
\begin{equation}
D^{2}(\rho ,\chi _{1,2}\left( \vec{p}\right) )=r_{o\alpha ^{\prime
}}^{2}+r_{o\alpha ^{\prime \prime }}^{2},  \label{c25}
\end{equation}%
and the optimal weights are
\begin{eqnarray}
p_{1} &=&\frac{1}{2}\left( 1+r_{o\alpha }\right) ,  \notag \\
p_{2} &=&1-p_{1},p_{3}=0.  \label{c26}
\end{eqnarray}

\textbf{Proof}. The mutually orthogonal $\left\vert \varphi
_{1}\right\rangle $ and $\left\vert \varphi _{2}\right\rangle $ mean that $%
\mathbf{r}_{1}^{T}\mathbf{r}_{2}=-1$.\ From $\mathbf{r}_{1}^{T}\mathbf{r}%
_{3}=\mathbf{r}_{2}^{T}\mathbf{r}_{3}=0$, we can get%
\begin{eqnarray}
\tilde{R}_{13} &=&\tilde{R}_{23}=-2,  \notag \\
\tilde{R}_{12} &=&0,d=4.
\end{eqnarray}%
Substituting these parameters into Theorem 2, one will obtain the
pseudo-probabilities for the problem  as%
\begin{eqnarray}
\tilde{p}_{1} &=&1-\tilde{p}_{2}-\tilde{p}_{3},  \notag \\
\tilde{p}_{2} &=&\frac{1}{2}\left( 1-r_{o\alpha }-r_{o\alpha ^{\prime
}}s_{3}\right) ,  \notag \\
\tilde{p}_{3} &=&r_{o\alpha ^{\prime }}s_{3}.
\end{eqnarray}%
The necessary and sufficient condition of restriction $\tilde{p}%
_{i}\geqslant 0$ holds for $i=1,2,3$ is%
\begin{equation}
r_{o\alpha }+r_{o\alpha ^{\prime }}\leq 1\mathfrak{,}s_{3}=1.
\end{equation}

If $s_{3}=1$\ and $r_{o\alpha }+r_{o\alpha ^{\prime }}>1$, the optimal
distance is given by
\begin{equation}
\min_{i<j}D(\rho ,\chi _{i,j}\left( \vec{p}\right) ),i,j,k=1,2,3.
\end{equation}%
By using Theorem 1, we can get%
\begin{eqnarray}
D^{2}(\rho ,\chi _{1,3}\left( \vec{p}\right) ) &=&r_{o\alpha ^{\prime \prime
}}^{2}+\frac{1}{2}\left( r_{o\alpha }+r_{o\alpha ^{\prime }}-1\right) ^{2},
\notag \\
D^{2}(\rho ,\chi _{1,2}\left( \vec{p}\right) ) &=&r_{o\alpha ^{\prime
}}^{2}+r_{o\alpha ^{\prime \prime }}^{2},  \label{c27}
\end{eqnarray}%
and%
\begin{eqnarray}
&&D^{2}(\rho ,\chi _{2,3}\left( \vec{p}\right) )  \notag \\
&=&\min \{D^{2}(\rho ,\chi _{2}\left( \vec{p}\right) ),D^{2}(\rho ,\chi
_{3}\left( \vec{p}\right) )\}  \notag \\
&=&\min \{1+\mathbf{r}_{o}^{T}\mathbf{r}_{o}+2r_{o\alpha },1+\mathbf{r}%
_{o}^{T}\mathbf{r}_{o}-2r_{o\alpha ^{\prime }}\}  \notag \\
&=&1+\mathbf{r}_{o}^{T}\mathbf{r}_{o}-2r_{o\alpha ^{\prime }}.  \label{c28}
\end{eqnarray}%
Next we compare the three quantities in Eq. (\ref{c27}) and Eq. (\ref{c28}).
We have%
\begin{eqnarray}
&&D^{2}(\rho ,\chi _{1,3}\left( \vec{p}\right) )-D^{2}(\rho ,\chi
_{1,2}\left( \vec{p}\right) )  \notag \\
&=&\frac{1}{2}\left( r_{o\alpha }+r_{o\alpha ^{\prime }}-1\right)
^{2}-r_{o\alpha ^{\prime }}^{2}  \notag \\
&=&\frac{1}{2}\left( r_{o\alpha }+r_{o\alpha ^{\prime }}-1+\sqrt{2}%
r_{o\alpha }\right) \left( r_{o\alpha }+r_{o\alpha ^{\prime }}-1-\sqrt{2}%
r_{o\alpha }\right)  \notag \\
&<&0.
\end{eqnarray}%
and%
\begin{eqnarray}
&&D^{2}(\rho ,\chi _{1,3}\left( \vec{p}\right) )-D^{2}(\rho ,\chi
_{2,3}\left( \vec{p}\right) )  \notag \\
&=&\frac{1}{2}\left( r_{o\alpha }+r_{o\alpha ^{\prime }}-1\right) ^{2}-1-%
\mathbf{r}_{o}^{T}\mathbf{r}_{o}+2r_{o\alpha ^{\prime }}  \notag \\
&=&-\frac{1}{2}\left( r_{o\alpha }-r_{o\alpha ^{\prime }}+1\right) ^{2}
\notag \\
&<&0.
\end{eqnarray}%
So $\min_{i<j}D(\rho ,\chi _{i,j}\left( \vec{p}\right) )$ can be realized by
the quantum state set $\{\left\vert \varphi _{1}\right\rangle ,\left\vert
\varphi _{3}\right\rangle \}$, and the optimal distance and weights can be
calculated by Eqs. (\ref{c23}) and (\ref{c24}), respectively.

If $s_{3}=-1$, we have%
\begin{eqnarray}
D^{2}(\rho ,\chi _{1,2}\left( \vec{p}\right) ) &=&r_{o\alpha ^{\prime
}}^{2}+r_{o\alpha ^{\prime \prime }}^{2},  \notag \\
D^{2}(\rho ,\chi _{2,3}\left( \vec{p}\right) ) &=&r_{o\alpha ^{\prime \prime
}}^{2}+\frac{1}{2}\left( r_{o\alpha }+r_{o\alpha ^{\prime }}+1\right) ^{2},
\label{c29}
\end{eqnarray}%
and%
\begin{eqnarray}
&&D^{2}(\rho ,\chi _{1,2}\left( \vec{p}\right) )-D^{2}(\rho ,\chi
_{2,3}\left( \vec{p}\right) )  \notag \\
&=&r_{o\alpha ^{\prime }}^{2}-\frac{1}{2}\left( r_{o\alpha }+r_{o\alpha
^{\prime }}+1\right) ^{2}  \notag \\
&=&\frac{1}{2}\left( \sqrt{2}r_{o\alpha ^{\prime }}+r_{o\alpha }+r_{o\alpha
^{\prime }}+1\right) \left( \sqrt{2}r_{o\alpha ^{\prime }}-r_{o\alpha
}-r_{o\alpha ^{\prime }}-1\right)  \notag \\
&<&0.  \label{c30}
\end{eqnarray}%
If $s_{3}=-1$ and $r_{o\alpha }+r_{o\alpha ^{\prime }}\leq 1$, we have%
\begin{equation}
D^{2}(\rho ,\chi _{1,3}\left( \vec{p}\right) )=r_{o\alpha ^{\prime \prime
}}^{2}+\frac{1}{2}\left( r_{o\alpha }-r_{o\alpha ^{\prime }}-1\right) ^{2},
\end{equation}%
and%
\begin{eqnarray}
&&D^{2}(\rho ,\chi _{1,2}\left( \vec{p}\right) )-D^{2}(\rho ,\chi
_{1,3}\left( \vec{p}\right) )  \notag \\
&=&r_{o\alpha ^{\prime }}^{2}-\frac{1}{2}\left( r_{o\alpha }-r_{o\alpha
^{\prime }}-1\right) ^{2}  \notag \\
&=&\frac{1}{2}\left( \sqrt{2}r_{o\alpha ^{\prime }}-r_{o\alpha ^{\prime
}}+r_{o\alpha }-1\right) \left( \sqrt{2}r_{o\alpha ^{\prime }}-r_{o\alpha
}+r_{o\alpha ^{\prime }}+1\right)  \notag \\
&<&0.  \label{c301}
\end{eqnarray}%
If $s_{3}=-1$ and $r_{o\alpha }+r_{o\alpha ^{\prime }}>1$, we have%
\begin{eqnarray}
&&D^{2}(\rho ,\chi _{1,3}\left( \vec{p}\right) )  \notag \\
&=&\min \{D^{2}(\rho ,\chi _{1}\left( \vec{p}\right) ),D^{2}(\rho ,\chi
_{3}\left( \vec{p}\right) )\}  \notag \\
&=&\min \{1+\mathbf{r}_{o}^{T}\mathbf{r}_{o}-2r_{o\alpha },1+\mathbf{r}%
_{o}^{T}\mathbf{r}_{o}+2r_{o\alpha ^{\prime }}\}  \notag \\
&=&1+\mathbf{r}_{o}^{T}\mathbf{r}_{o}-2r_{o\alpha },  \label{c291}
\end{eqnarray}%
and%
\begin{eqnarray}
&&D^{2}(\rho ,\chi _{1,2}\left( \vec{p}\right) )-D^{2}(\rho ,\chi
_{1,3}\left( \vec{p}\right) )  \notag \\
&=&r_{o\alpha ^{\prime }}^{2}-1-\mathbf{r}_{o}^{T}\mathbf{r}_{o}+2r_{o\alpha
}  \notag \\
&=&-\left( 1-r_{o\alpha }\right) ^{2}<0.  \label{c302}
\end{eqnarray}%
So $\min_{i<j}D(\rho ,\chi _{i,j}\left( \vec{p}\right) )$ can be realized by
the quantum state set $\{\left\vert \varphi _{1}\right\rangle ,\left\vert
\varphi _{2}\right\rangle \}$, and the optimal distance and weights can be
calculated by Eqs. (\ref{c25}) and (\ref{c26}), respectively. The proof is
completed. \hfill $\blacksquare $

\textbf{Case 2}.- If $\left\vert \varphi _{1}\right\rangle =\left\vert
x\right\rangle _{s_{x}}$, $\left\vert \varphi _{2}\right\rangle =\left\vert
y\right\rangle _{s_{y}}$ and $\left\vert \varphi _{3}\right\rangle
=\left\vert z\right\rangle _{s_{z}}$, the pseudo-probabilities for the
problem can be first given by%
\begin{eqnarray}
\tilde{p}_{x} &=&\frac{1}{3}\left(
1+2r_{ox}s_{x}-r_{oy}s_{y}-r_{oz}s_{z}\right) ,  \notag \\
\tilde{p}_{y} &=&\frac{1}{3}\left(
1+2r_{oy}s_{y}-r_{ox}s_{x}-r_{oz}s_{z}\right) ,  \notag \\
\tilde{p}_{z} &=&1-\tilde{p}_{x}-\tilde{p}_{y},
\end{eqnarray}%
If $\tilde{p}_{i}\geqslant 0$ holds for $i=x,y,z$, the optimal weights $%
p_{i}=\tilde{p}_{i}$, and the optimal distance are given by%
\begin{equation}
D(\rho ,\chi _{x,y,z}\left( \vec{p}\right) )=\frac{\sqrt{3}}{3}\left(
1-2r_{ox}s_{x}-r_{oy}s_{y}-r_{oz}s_{z}\right) .  \label{c311}
\end{equation}%
If $s_{x}=s_{y}=s_{z}$ and $\tilde{p}_{\alpha }<0$ with $\left\{ \alpha \neq
\alpha ^{\prime }\neq \alpha ^{\prime \prime }|\alpha ,\alpha ^{\prime
},\alpha ^{\prime \prime }=x,y,z\right\} $, the optimal distance is given by%
\begin{equation}
D^{2}(\rho ,\chi _{\alpha ^{\prime },\alpha ^{\prime \prime }}\left( \vec{p}%
\right) )=r_{o\alpha }^{2}+\frac{1}{2}\left( r_{o\alpha ^{\prime
}}+r_{o\alpha ^{\prime \prime }}-s_{x}\right) ^{2},  \label{c33}
\end{equation}%
and the optimal weights are given by
\begin{eqnarray}
p_{\alpha ^{\prime }} &=&\frac{1}{2}\left[ 1+s_{x}\left( r_{o\alpha ^{\prime
}}-r_{o\alpha ^{\prime \prime }}\right) \right] ,  \notag \\
p_{\alpha ^{\prime \prime }} &=&1-p_{\alpha ^{\prime }},p_{\alpha }=0.
\label{c34}
\end{eqnarray}%
If $s_{\alpha }=1$, $s_{\alpha ^{\prime }}=s_{\alpha ^{\prime \prime }}=-1$,
$p_{\alpha ^{\prime }}<0$ and $r_{o\alpha }+r_{o\alpha ^{\prime \prime
}}\leq 1$, the optimal distance is given by%
\begin{equation}
D(\rho ,\chi _{\alpha ,\alpha ^{\prime \prime }}\left( \vec{p}\right)
)=r_{o\alpha ^{\prime }}^{2}+\frac{1}{2}\left( r_{o\alpha }-r_{o\alpha
^{\prime \prime }}-1\right) ^{2},  \label{c35}
\end{equation}%
and the optimal weights are
\begin{eqnarray}
p_{\alpha } &=&\frac{1}{2}\left( 1+r_{o\alpha }+r_{o\alpha ^{\prime \prime
}}\right) ,  \notag \\
p_{\alpha ^{\prime \prime }} &=&1-p_{\alpha },p_{\alpha ^{\prime }}=0.
\label{c36}
\end{eqnarray}%
If $s_{\alpha }=1$, $s_{\alpha ^{\prime }}=s_{\alpha ^{\prime \prime }}=-1$,
$p_{\alpha ^{\prime }}<0$ and $r_{o\alpha }+r_{o\alpha ^{\prime \prime }}>1$%
, the optimal distance is given by%
\begin{equation}
D(\rho ,\chi _{\alpha }\left( \vec{p}\right) )=1+\mathbf{r}_{o}^{T}\mathbf{r}%
_{o}-2r_{o\alpha },  \label{c37}
\end{equation}%
and the optimal weights are
\begin{equation}
p_{\alpha }=1,p_{\alpha ^{\prime }}=p_{\alpha ^{\prime \prime }}=0.
\label{c38}
\end{equation}%
If $s_{\alpha }=s_{\alpha ^{\prime }}=1$, $s_{\alpha ^{\prime \prime }}=-1$\
and $\tilde{p}_{\alpha ^{\prime \prime }}<0$, the optimal distance is given
by%
\begin{equation}
D(\rho ,\chi _{\alpha ,\alpha ^{\prime }}\left( \vec{p}\right) )=r_{o\alpha
^{\prime \prime }}^{2}+\frac{1}{2}\left( r_{o\alpha }+r_{o\alpha ^{\prime
}}-1\right) ^{2},  \label{c39}
\end{equation}%
and the optimal weights are
\begin{eqnarray}
p_{\alpha } &=&\frac{1}{2}\left( 1+r_{o\alpha }-r_{o\alpha ^{\prime
}}\right) ,  \notag \\
p_{\alpha ^{\prime }} &=&1-p_{\alpha },p_{\alpha ^{\prime \prime }}=0.
\label{c399}
\end{eqnarray}

\textbf{Proof}. For these three special quantum states, we have%
\begin{equation}
\mathbf{r}_{i}^{T}\mathbf{r}_{j}=0,\tilde{R}_{ij}=-1,d=3,
\end{equation}%
with $i,j=x,y,z$. Substitute these parameters into Theorem 2, one will
obtain
\begin{eqnarray}
\tilde{p}_{x} &=&\frac{1}{3}\left(
1+2r_{ox}s_{x}-r_{oy}s_{y}-r_{oz}s_{z}\right) ,  \notag \\
\tilde{p}_{y} &=&\frac{1}{3}\left(
1+2r_{oy}s_{y}-r_{ox}s_{x}-r_{oz}s_{z}\right) ,  \notag \\
\tilde{p}_{z} &=&1-\tilde{p}_{x}-\tilde{p}_{y},
\end{eqnarray}%
If $\tilde{p}_{i}\geqslant 0$ holds for $i=x,y,z$, the optimal weights are
given by $p_{i}=\tilde{p}_{i}$ and the optimal distance%
\begin{equation}
D(\rho ,\chi _{x,y,z}\left( \vec{p}\right) )=\frac{\sqrt{3}}{3}\left(
1-2r_{ox}s_{x}-r_{oy}s_{y}-r_{oz}s_{z}\right) .
\end{equation}
Otherwise, we consider the following three cases.

\textit{Case 2.1}\textbf{.- }$s_{x}=s_{y}=s_{z}$. If $\tilde{p}_{\alpha }<0$
with $\left\{ \alpha \neq \alpha ^{\prime }\neq \alpha ^{\prime \prime
}|\alpha ,\alpha ^{\prime },\alpha ^{\prime \prime }=x,y,z\right\} $, the
optimal distance is given by
\begin{equation}
\min_{i<j}D(\rho ,\chi _{i,j}\left( \vec{p}\right) ),i,j=x,y,z.
\end{equation}%
From $\tilde{p}_{\alpha }<0$, we can get%
\begin{equation*}
1+s_{x}\left( 2r_{o\alpha }-r_{o\alpha ^{\prime }}-r_{o\alpha ^{\prime
\prime }}\right) <0.
\end{equation*}%
Using Theorem 1, we can get%
\begin{equation}
D^{2}(\rho ,\chi _{\alpha ,\alpha ^{\prime }}\left( \vec{p}\right)
)=r_{o\alpha ^{\prime \prime }}^{2}+\frac{1}{2}\left( r_{o\alpha
}+r_{o\alpha ^{\prime }}-s_{x}\right) ^{2}.  \label{F2}
\end{equation}%
Next we compare the above three combinations. We
have%
\begin{eqnarray}
&&D^{2}(\rho ,\chi _{\alpha ^{\prime },\alpha ^{\prime \prime }}\left( \vec{p%
}\right) )-D^{2}(\rho ,\chi _{\alpha ,\alpha ^{\prime }}\left( \vec{p}%
\right) )  \notag \\
&=&r_{o\alpha }^{2}+\frac{1}{2}\left( r_{o\alpha ^{\prime }}+r_{o\alpha
^{\prime \prime }}-s_{x}\right) ^{2}-r_{o\alpha ^{\prime \prime }}^{2}-\frac{%
1}{2}\left( r_{o\alpha }+r_{o\alpha ^{\prime }}-s_{x}\right) ^{2}  \notag \\
&=&\frac{1}{2}\left( r_{o\alpha ^{\prime \prime }}-r_{o\alpha }\right)
\left( 2r_{o\alpha ^{\prime }}-2s_{x}-r_{o\alpha }-r_{o\alpha ^{\prime
\prime }}\right)  \notag \\
&<&0,
\end{eqnarray}%
and%
\begin{eqnarray}
&&D^{2}(\rho ,\chi _{\alpha ^{\prime },\alpha ^{\prime \prime }}\left( \vec{p%
}\right) )-D^{2}(\rho ,\chi _{\alpha ,\alpha ^{\prime \prime }}\left( \vec{p}%
\right) )  \notag \\
&=&r_{o\alpha }^{2}+\frac{1}{2}\left( r_{o\alpha ^{\prime }}+r_{o\alpha
^{\prime \prime }}-s_{x}\right) ^{2}-r_{o\alpha ^{\prime }}^{2}-\frac{1}{2}%
\left( r_{o\alpha }+r_{o\alpha ^{\prime \prime }}-s_{x}\right) ^{2}  \notag
\\
&=&\frac{1}{2}\left( r_{o\alpha ^{\prime }}-r_{o\alpha }\right) \left(
2r_{o\alpha ^{\prime \prime }}-2s_{x}-r_{o\alpha }-r_{o\alpha ^{\prime
}}\right)  \notag \\
&<&0.
\end{eqnarray}%
So $\min_{i<j}D(\rho ,\chi _{i,j}\left( \vec{p}\right) )$ can be realized by
the quantum state set $\{\left\vert \alpha ^{\prime }\right\rangle
,\left\vert \alpha ^{\prime \prime }\right\rangle \}$, and the optimal
distance and the weights can be calculated as Eqs. (\ref{c33}) and (\ref{c34}),
respectively.

\textit{Case 2.2}\textbf{.- }$s_{\alpha }=1$, $s_{\alpha ^{\prime
}}=s_{\alpha ^{\prime \prime }}=-1$. We can find that $\tilde{p}_{\alpha }=%
\frac{1}{3}\left( 1+2r_{o\alpha }+r_{o\alpha ^{\prime }}+r_{o\alpha ^{\prime
\prime }}\right) >0$. If $\tilde{p}_{\alpha ^{\prime }}<0$ with $\left\{
\alpha \neq \alpha ^{\prime }\neq \alpha ^{\prime \prime }|\alpha ,\alpha
^{\prime },\alpha ^{\prime \prime }=x,y,z\right\} $, the optimal distance is
given by
\begin{equation}
\min_{i<j}D(\rho ,\chi _{i,j}\left( \vec{p}\right) ),i,j=x,y,z.
\end{equation}%
From $\tilde{p}_{\alpha ^{\prime }}<0$, we can get%
\begin{equation}
1+r_{o\alpha ^{\prime \prime }}-2r_{o\alpha ^{\prime }}-r_{o\alpha }<0.
\label{c221}
\end{equation}%
By using Theorem 1, we can get%
\begin{equation}
D^{2}(\rho ,\chi _{\alpha ^{\prime },\alpha ^{\prime \prime }}\left( \vec{p}%
\right) )=r_{o\alpha }^{2}+\frac{1}{2}\left( r_{o\alpha ^{\prime
}}+r_{o\alpha ^{\prime \prime }}+1\right) ^{2}.
\end{equation}

\textit{Case 2.2.1}\textbf{.- }$r_{o\alpha }+r_{o\alpha ^{\prime \prime
}}\leq 1$. Based on Theorem 1, we can get%
\begin{equation}
D^{2}(\rho ,\chi _{\alpha ,\alpha ^{\prime \prime }}\left( \vec{p}\right)
)=r_{o\alpha ^{\prime }}^{2}+\frac{1}{2}\left( r_{o\alpha }-r_{o\alpha
^{\prime \prime }}-1\right) ^{2}.
\end{equation}%
Thus we have%
\begin{eqnarray}
&&D^{2}(\rho ,\chi _{\alpha ,\alpha ^{\prime \prime }}\left( \vec{p}\right)
)-D^{2}(\rho ,\chi _{\alpha ,\alpha ^{\prime }}\left( \vec{p}\right) )
\notag \\
&=&\frac{1}{2}\left( r_{o\alpha ^{\prime \prime }}-r_{o\alpha ^{\prime
}}\right) \left( 2-2r_{o\alpha }-r_{o\alpha ^{\prime }}-r_{o\alpha ^{\prime
\prime }}\right)  \notag \\
&<&0.
\end{eqnarray}%
If $r_{o\alpha }+r_{o\alpha ^{\prime }}\leq 1$, we have%
\begin{equation}
D^{2}(\rho ,\chi _{\alpha ,\alpha ^{\prime }}\left( \vec{p}\right)
)=r_{o\alpha ^{\prime \prime }}^{2}+\frac{1}{2}\left( r_{o\alpha
}-r_{o\alpha ^{\prime }}-1\right) ^{2},
\end{equation}%
and%
\begin{eqnarray}
&&D^{2}(\rho ,\chi _{\alpha ,\alpha ^{\prime \prime }}\left( \vec{p}\right)
)-D^{2}(\rho ,\chi _{\alpha ,\alpha ^{\prime }}\left( \vec{p}\right) )
\notag \\
&=&\frac{1}{2}\left( r_{o\alpha ^{\prime \prime }}-r_{o\alpha ^{\prime
}}\right) \left( 2-2r_{o\alpha }-r_{o\alpha ^{\prime }}-r_{o\alpha ^{\prime
\prime }}\right)  \notag \\
&<&0.
\end{eqnarray}%
If $r_{o\alpha }+r_{o\alpha ^{\prime }}>1$, we have%
\begin{equation}
D^{2}(\rho ,\chi _{\alpha ,\alpha ^{\prime }}\left( \vec{p}\right)
)=D^{2}(\rho ,\chi _{\alpha }\left( \vec{p}\right) ),
\end{equation}%
and%
\begin{equation}
D^{2}(\rho ,\chi _{\alpha ,\alpha ^{\prime \prime }}\left( \vec{p}\right)
)<D^{2}(\rho ,\chi _{\alpha }\left( \vec{p}\right) ).
\end{equation}%
So $\min_{i<j}D(\rho ,\chi _{i,j}\left( \vec{p}\right) )$ can be realized by
the quantum state set $\{\left\vert \alpha \right\rangle ,\left\vert \alpha
^{\prime \prime }\right\rangle \}$, and the optimal distance and the weights can
be calculated as Eqs. (\ref{c35}) and (\ref{c36}), respectively.

\textit{Case 2.2.2}\textbf{.- }$r_{o\alpha }+r_{o\alpha ^{\prime \prime }}>1$%
. Combined with Eq. (\ref{c221}), we can get $r_{o\alpha }+r_{o\alpha
^{\prime }}>1$ and $r_{o\alpha ^{\prime }}>r_{o\alpha ^{\prime \prime }}$.
By using Theorem 1, we can get%
\begin{eqnarray}
D^{2}(\rho ,\chi _{\alpha ,\alpha ^{\prime }}\left( \vec{p}\right) )
&=&D^{2}(\rho ,\chi _{\alpha ,\alpha ^{\prime \prime }}\left( \vec{p}\right)
)  \notag \\
&=&D^{2}(\rho ,\chi _{\alpha }\left( \vec{p}\right) )  \notag \\
&=&1+\mathbf{r}_{o}^{T}\mathbf{r}_{o}-2r_{o\alpha },
\end{eqnarray}%
and%
\begin{eqnarray}
&&D^{2}(\rho ,\chi _{\alpha ,\alpha ^{\prime }}\left( \vec{p}\right)
)-D^{2}(\rho ,\chi _{\alpha ^{\prime },\alpha ^{\prime \prime }}\left( \vec{p%
}\right) )  \notag \\
&=&\frac{1}{2}\left[ 4\left( 1-r_{o\alpha }-r_{o\alpha ^{\prime }}\right)
+\left( r_{o\alpha ^{\prime }}-r_{o\alpha ^{\prime \prime }}\right) \left(
2+r_{o\alpha ^{\prime }}-r_{o\alpha ^{\prime \prime }}\right) -3\right]
\notag \\
&<&0.
\end{eqnarray}%
So $\min_{i<j}D(\rho ,\chi _{i,j}\left( \vec{p}\right) )$ can be realized by
the quantum state set $\{\left\vert \alpha \right\rangle \}$. The
optimal distance and the weights can be calculated as Eqs. (\ref{c37}) and (\ref%
{c38}), respectively.

\textit{Case 2.3}\textbf{.- }$s_{\alpha }=s_{\alpha ^{\prime }}=1$, $%
s_{\alpha ^{\prime \prime }}=-1$. We can find that $\tilde{p}_{\alpha }=%
\frac{1}{3}\left( 1+2r_{o\alpha }+r_{o\alpha `}-r_{o\alpha ^{\prime \prime
}}\right) >0$ and $\tilde{p}_{\alpha ^{\prime }}=\frac{1}{3}\left(
1+2r_{o\alpha ^{\prime }}+r_{o\alpha }-r_{o\alpha ^{\prime \prime }}\right)
>0$. If $\tilde{p}_{\alpha ^{\prime \prime }}<0$, the optimal distance is
given by
\begin{equation}
\min_{i<j}D(\rho ,\chi _{i,j}\left( \vec{p}\right) ),i,j=x,y,z.
\end{equation}%
By using Theorem 1, we can get%
\begin{equation}
D^{2}(\rho ,\chi _{\alpha ,\alpha ^{\prime }}\left( \vec{p}\right)
)=r_{o\alpha ^{\prime \prime }}^{2}+\frac{1}{2}\left( r_{o\alpha
}-r_{o\alpha ^{\prime }}-1\right) ^{2}.
\end{equation}%
If $r_{o\alpha }+r_{o\alpha ^{\prime \prime }}\leq 1$, we have%
\begin{equation}
D^{2}(\rho ,\chi _{\alpha ,\alpha ^{\prime \prime }}\left( \vec{p}\right)
)=r_{o\alpha ^{\prime }}^{2}+\frac{1}{2}\left( r_{o\alpha }-r_{o\alpha
^{\prime \prime }}-1\right) ^{2},  \label{c231}
\end{equation}%
and%
\begin{eqnarray}
&&D(\rho ,\chi _{\alpha ,\alpha ^{\prime }}\left( \vec{p}\right) )-D(\rho
,\chi _{\alpha ,\alpha ^{\prime \prime }}\left( \vec{p}\right) )  \notag \\
&=&\frac{1}{2}\left( r_{o\alpha ^{\prime }}+r_{o\alpha ^{\prime \prime
}}\right) \left( r_{o\alpha ^{\prime \prime }}+2r_{o\alpha }-2-r_{o\alpha
^{\prime }}\right)  \notag \\
&<&0.
\end{eqnarray}%
If $r_{o\alpha }+r_{o\alpha ^{\prime \prime }}>1$, we have%
\begin{equation}
D(\rho ,\chi _{\alpha ,\alpha ^{\prime \prime }}\left( \vec{p}\right)
)=D(\rho ,\chi _{\alpha }\left( \vec{p}\right) )=1+\mathbf{r}_{o}^{T}\mathbf{%
r}_{o}-2r_{o\alpha },
\end{equation}%
and%
\begin{equation}
D(\rho ,\chi _{\alpha ,\alpha ^{\prime }}\left( \vec{p}\right) )\leq D(\rho
,\chi _{\alpha }\left( \vec{p}\right) ).  \label{c232}
\end{equation}%
Through a similar process from Eq. (\ref{c231}) to Eq. (\ref{c232}), we can
get%
\begin{equation}
D(\rho ,\chi _{\alpha ,\alpha ^{\prime }}\left( \vec{p}\right) )\leq D(\rho
,\chi _{\alpha ^{\prime },\alpha ^{\prime \prime }}\left( \vec{p}\right) ).
\end{equation}%
To sum up, we can conclude that if $s_{\alpha }=s_{\alpha ^{\prime }}=1$, $%
s_{\alpha ^{\prime \prime }}=-1$, then $\min_{i<j}D(\rho ,\chi _{i,j}\left(
\vec{p}\right) )$ can be realized by the quantum state set $\{\left\vert
\alpha \right\rangle ,\left\vert \alpha ^{\prime }\right\rangle \}$. The
optimal distance and the weights can be calculated as Eqs. (\ref{c39}) and (\ref%
{c399}), respectively. The proof is completed.\hfill $\blacksquare $

For $N=4$, we choose four of the six quantum states to solve the convex
optimization problem.

\textbf{Case 3}.- We set $\left\vert \varphi _{1\pm }\right\rangle
=\left\vert \alpha \right\rangle _{\pm }$ and $\left\vert \varphi _{2\pm
}\right\rangle =\left\vert \alpha ^{\prime }\right\rangle _{\pm }$ with $%
\left\{ \alpha \neq \alpha ^{\prime }\neq \alpha ^{\prime \prime }|\alpha
,\alpha ^{\prime },\alpha ^{\prime \prime }=x,y,z\right\} $. If $r_{o\alpha
}+r_{o\alpha ^{\prime }}\leq 1$, one can calculate the optimal distance as%
\begin{equation}
D(\rho ,\chi _{1,2,3}\left( \vec{p}\right) )=r_{o\alpha ^{\prime \prime }},
\end{equation}%
and the optimal weights as
\begin{eqnarray}
p_{1+} &=&r_{o\alpha },p_{1-}=0,  \notag \\
p_{2-} &=&\frac{1}{2}\left( 1-r_{o\alpha ^{\prime }}-r_{o\alpha }\right) ,
\notag \\
p_{2+} &=&1-p_{1+}-p_{2-}.  \label{CC31}
\end{eqnarray}%
or%
\begin{eqnarray}
p_{2+} &=&r_{o\alpha ^{\prime }},p_{2-}=0,  \notag \\
p_{1-} &=&\frac{1}{2}\left( 1-r_{o\alpha }-r_{o\alpha ^{\prime }}\right) ,
\notag \\
p_{1+} &=&1-p_{1-}-p_{2+};  \label{CC32}
\end{eqnarray}%
If $r_{o\alpha }+r_{o\alpha ^{\prime }}>1$, the optimal distance is given by%
\begin{equation}
D^{2}(\rho ,\chi _{1,2}\left( \vec{p}\right) )=r_{o\alpha ^{\prime \prime
}}^{2}+\frac{1}{2}\left( r_{o\alpha }+r_{o\alpha ^{\prime }}-1\right) ^{2},
\label{CC33}
\end{equation}%
and the optimal weights are given by%
\begin{eqnarray}
p_{1+} &=&\frac{1}{2}\left( 1+r_{o\alpha }-r_{o\alpha ^{\prime }}\right) ,
\notag \\
p_{2+} &=&1-p_{1+},p_{1-}=p_{2-}=0.  \label{CC34}
\end{eqnarray}

\textbf{Proof}. If $\left\vert \varphi _{1\pm }\right\rangle =\left\vert
\alpha \right\rangle _{\pm }$ and $\left\vert \varphi _{2\pm }\right\rangle
=\left\vert \alpha ^{\prime }\right\rangle _{\pm }$ with $\left\{ \alpha
\neq \alpha ^{\prime }\neq \alpha ^{\prime \prime }|\alpha ,\alpha ^{\prime
},\alpha ^{\prime \prime }=x,y,z\right\} $, the rank of matrix A is equal to
3. According to Theorem 4, the optimal distance is given by
\begin{equation}
\min_{i<j<k}D(\rho ,\chi _{i,j,k}\left( \vec{p}\right) ),i,j,k=1,2,3,4.
\label{T61}
\end{equation}%
According to \textbf{Case 1}, if we choose three quantum states from $%
\{\left\vert \varphi _{1\pm }\right\rangle ,\left\vert \varphi _{2\pm
}\right\rangle \}$, we can only choose $\{\left\vert \varphi
_{1+}\right\rangle ,\left\vert \varphi _{2\pm }\right\rangle \}$ and $%
\{\left\vert \varphi _{1\pm }\right\rangle ,\left\vert \varphi
_{2+}\right\rangle \}$. And the corresponding pseudo-probabilities are Eqs. (%
\ref{CC31}) and (\ref{CC32}). The necessary and sufficient condition of
restriction $\tilde{p}_{i}\geqslant 0$ holds for $i=1,2,3$ is%
\begin{equation}
r_{o\alpha ^{\prime }}+r_{o\alpha }\leq 1.
\end{equation}%
If $r_{o\alpha }+r_{o\alpha ^{\prime }}>1$, the optimal distance is given by
\begin{equation}
\min_{i<j}D(\rho ,\chi _{i,j}\left( \vec{p}\right) ),i,j,k=1,2,3,4.
\end{equation}%
According to \textbf{Case 1}, we can find $\min_{i<j}D(\rho ,\chi
_{i,j}\left( \vec{p}\right) )$ can be realized by the quantum state set $%
\{\left\vert \varphi _{1+}\right\rangle ,\left\vert \varphi
_{2+}\right\rangle \}$. And the optimal distance and weights can be
calculated as Eqs. (\ref{CC33}) and (\ref{CC34}), respectively. The proof is
completed.\hfill $\blacksquare $

\textbf{Case 4}.- We set $\left\vert \varphi _{1}\right\rangle =\left\vert
\alpha \right\rangle _{1}$, $\left\vert \varphi _{2}\right\rangle
=\left\vert \alpha \right\rangle _{-1}$, $\left\vert \varphi
_{3}\right\rangle =\left\vert \alpha ^{\prime }\right\rangle _{s_{3}}$ and $%
\left\vert \varphi _{4}\right\rangle =\left\vert \alpha ^{\prime \prime
}\right\rangle _{s_{4}}$ with $\left\{ \alpha \neq \alpha ^{\prime }\neq
\alpha ^{\prime \prime }|\alpha ,\alpha ^{\prime },\alpha ^{\prime \prime
}=x,y,z\right\} $. If%
\begin{eqnarray}
&&s_{3}=1,s_{4}=1,  \notag \\
&&r_{ox}+r_{oy}+r_{oz}\leq 1,
\end{eqnarray}%
the objective state $\rho $ can be exactly written as the convex sum of the
4 pure states. And the corresponding probabilities are given by
\begin{eqnarray}
\tilde{p}_{1} &=&\frac{1}{2}\left( 1+r_{o\alpha }-r_{o\alpha ^{\prime
}}-r_{o\alpha ^{\prime \prime }}\right)  \notag \\
\tilde{p}_{2} &=&\frac{1}{2}\left( 1-r_{o\alpha }-r_{o\alpha ^{\prime
}}-r_{o\alpha ^{\prime \prime }}\right)  \notag \\
\tilde{p}_{3} &=&r_{o\alpha ^{\prime }},\tilde{p}_{4}=r_{o\alpha ^{\prime
\prime }}.
\end{eqnarray}%
Otherwise, the optimal approximation is solved based on Case 2 as
\begin{equation}
\min_{i<j<k}D(\rho ,\chi _{i,j,k}\left( \vec{p}\right) ),i,j,k=1,2,3,4.
\label{T63}
\end{equation}

\textbf{Proof}. The rank of matrix $A$ is equal to 4. For Pauli matrix, we
have%
\begin{eqnarray}
\mathbf{\vec{r}}_{x_{\pm }} &=&\left(
\begin{array}{ccc}
\pm 1 & 0 & 0%
\end{array}%
\right) ^{T},  \notag \\
\mathbf{\vec{r}}_{y_{\pm }} &=&\left(
\begin{array}{ccc}
0 & \pm 1 & 0%
\end{array}%
\right) ^{T},  \notag \\
\mathbf{\vec{r}}_{z_{\pm }} &=&\left(
\begin{array}{ccc}
0 & 0 & \pm 1%
\end{array}%
\right) ^{T}.
\end{eqnarray}%
Substituting these parameters into Theorem 3, one will obtain $A$. For
example, if $\left\vert \varphi _{1}\right\rangle =\left\vert x\right\rangle
_{1}$, $\left\vert \varphi _{2}\right\rangle =\left\vert x\right\rangle
_{-1} $, $\left\vert \varphi _{3}\right\rangle =\left\vert y\right\rangle
_{1}$ and $\left\vert \varphi _{4}\right\rangle =\left\vert z\right\rangle
_{-1}$, we have%
\begin{equation}
A=\left(
\begin{array}{cccc}
1 & -1 & 0 & 0 \\
0 & 0 & 1 & 0 \\
0 & 0 & 0 & -1 \\
1 & 1 & 1 & 1%
\end{array}%
\right) ,
\end{equation}%
and%
\begin{equation}
A^{-1}=\frac{1}{2}\left(
\begin{array}{cccc}
1 & -1 & 1 & 1 \\
-1 & -1 & 1 & 1 \\
0 & 2 & 0 & 0 \\
0 & 0 & -2 & 0%
\end{array}%
\right) .
\end{equation}%
By
\begin{equation}
\mathbf{\tilde{p}=}A^{-1}\tilde{r}
\end{equation}%
in Theorem 3, we can get%
\begin{eqnarray}
\tilde{p}_{1} &=&\frac{1}{2}\left( 1+r_{o\alpha }-r_{o\alpha ^{\prime
}}s_{3}-r_{o\alpha ^{\prime \prime }}s_{4}\right)  \notag \\
\tilde{p}_{2} &=&\frac{1}{2}\left( 1-r_{o\alpha }-r_{o\alpha ^{\prime
}}s_{3}-r_{o\alpha ^{\prime \prime }s_{4}}\right)  \notag \\
\tilde{p}_{3} &=&r_{o\alpha ^{\prime }}s_{3},\tilde{p}_{4}=r_{o\alpha
^{\prime \prime }}s_{4}.
\end{eqnarray}%
The necessary and sufficient condition of restriction $\tilde{p}%
_{i}\geqslant 0$ holds for $i=1,2,3,4$ is%
\begin{eqnarray}
&&s_{3}=1,s_{4}=1,  \notag \\
&&r_{ox}+r_{oy}+r_{oz}\leq 1.
\end{eqnarray}%
Otherwise, the optimal approximation is solved based on Case 2 as
\begin{equation}
\min_{i<j<k}D(\rho ,\chi _{i,j,k}\left( \vec{p}\right) ),i,j,k=1,2,3,4.
\end{equation}%
The proof is completed. \hfill $\blacksquare $

For $N=6$, we choose all six quantum states $\left\{ \left\vert
x\right\rangle _{s},\left\vert y\right\rangle _{s},\left\vert z\right\rangle
_{s}|s=\pm 1\right\} $ to solve the convex optimization problem. According
to theorem 5, the problem is equivalent to selecting four most suitable
eigenstates from six eigenstates to find the best. The optimal distance is
given by%
\begin{equation}
\min_{i<j<k<l}D(\rho ,\chi _{i,j,k,l}\left( \vec{p}\right)
),i,j,k,l=1,2,3,4,5,6.
\end{equation}%
which as well as the corresponding weights can be solved by \textbf{Case 3}
and \textbf{Case 4}.

\end{document}